\documentclass{aa}  
\usepackage[table]{xcolor}

\usepackage{graphicx}
\usepackage{txfonts}
\usepackage[colorlinks=True,linkcolor=blue,citecolor=blue]{hyperref}
\usepackage{url}
\usepackage{mathtools}
 \usepackage{booktabs}
\usepackage{multirow}
 \usepackage{placeins}  
\usepackage{orcidlink}

\usepackage{etoolbox}
\makeatletter
\patchcmd{\appendix}{\clearpage}{}{}{}
\makeatother
\begin{document}

   \title{There Is More to Outshining: 2D Dust Effects on Stellar Mass Estimates at $3\leq z<9$ with JWST in the JADES Field
   }
\author{M. Hamed\inst{1}\orcidlink{0000-0001-9626-9642}
\and
{P. G. P\'erez-Gonz\'alez\inst{1}}
\and 
M. Annunziatella\inst{2}
\and
L. Colina\inst{1}
\and
I. Shivaei\inst{1}
\and
M. Perna\inst{1}
\and
A.J. Bunker\inst{3}
\and
K. Ma{\l}ek\inst{4, 5}
\and
S. Arribas\inst{1}
\and
J. \'Alvarez-M\'arquez\inst{1}
\and
C.N.A. Willmer\inst{6}
\and
H. \"Ubler\inst{7}
\and
R. Bhatawdekar\inst{8}
\and
J. Chevallard\inst{3}
\and
E. Curtis-Lake\inst{9}
\and
Z. Ji\inst{6}
\and
P. Rinaldi\inst{10}
\and 
C.C. Williams\inst{11}
}

\institute{Centro de Astrobiolog\'ia (CAB), CSIC-INTA, Ctra. de Ajalvir km 4, Torrej\'on de Ardoz, E-28850, Madrid, Spain 
\and
INAF -- IASF Milano, Via A. Corti 12, 20133 Milano, Italy
\and 
Department of Physics, University of Oxford, Denys Wilkinson Building, Keble Road, Oxford OX1 3RH, UK
\and
National Centre for Nuclear Research, Pasteura 7, 02-093, Warsaw, Poland
\and 
Aix Marseille Univ. CNRS, CNES, LAM, Marseille, France
\and
Steward Observatory, University of Arizona, 933 N. Cherry Avenue, Tucson, AZ 85721, USA
\and
Max-Planck-Institut f\"ur extraterrestrische Physik (MPE), Gie{\ss}enbachstra{\ss}e 1, 85748 Garching, Germany
\and
European Space Agency (ESA), European Space Astronomy Centre (ESAC), Camino Bajo del Castillo s/n, 28692 Villanueva de la Cañada, Madrid, Spain
\and
Centre for Astrophysics Research, Department of Physics, Astronomy and Mathematics, University of Hertfordshire, Hatfield AL10 9AB, UK
\and 
Space Telescope Science Institute, 3700 San Martin Drive, Baltimore, Maryland 21218, USA
\and
NSF National Optical-Infrared Astronomy Research Laboratory, 950 North Cherry Avenue, Tucson, AZ 85719, USA
}


   \date{}

\abstract{}{
Dust attenuation modifies the observed spectral energy distribution (SED), leading to biases in the physical parameters inferred from integrated SED fitting linked to degeneracies (e.g., between dust content and age). As spatially resolved SED modeling becomes feasible for large high-redshift samples, it is increasingly important to assess how dust attenuation affects resolved mass estimates. We evaluate the impact of dust attenuation on stellar mass estimates derived from integrating spatially resolved SED fitting results.}
{We perform spatially resolved and integrated SED fitting on a sample of 3\,408 galaxies at $3 \leq z < 9$ from the Great Observatories Origins Deep Survey (GOODS) South field, combining deep NIRCam from the JWST Advanced Deep Extragalactic Survey (JADES) and HST/ACS imaging from GOODS and the Cosmic Assembly Near-infrared Deep Extragalactic Legacy Survey (CANDELS). We compare galaxy-integrated properties derived from fitting the summed SED with those obtained from spatially resolved SED modeling. Using a two-component dust attenuation model with a variable slope, we investigate how the dust attenuation slope, $A_V$, and stellar population properties contribute to discrepancies in the resulting stellar mass estimates.}
{Resolved stellar masses are systematically higher than integrated estimates, with a median offset of $\Delta\log \mathrm{M}_\star = +0.24$ dex. This offset exhibits a strong mass dependence: +0.34 dex at low masses ($\log \mathrm{M}_\star/\mathrm{M_{\odot}}  < 8$), decreasing to +0.22 dex at intermediate masses ($\log\mathrm{M}_\star/\mathrm{M_{\odot}} \sim 8.5$), and approaching agreement (+0.04 dex) at the highest masses ($\log \mathrm{M}_\star/\mathrm{M_{\odot}} > 10$). This demonstrates that outshining, where young stellar populations dominate the integrated light, disproportionately affects low-mass galaxies and leads to mass underestimation by factors of $\sim$2. Resolved analyses recover higher dust attenuations ($\Delta A_V \approx +0.08$~mag), lower birth cloud (BC) fractions ($\Delta\mu  \approx -0.28$, with $\mu \equiv \frac{A_{V,\mathrm{ISM}}}{A_{V,\mathrm{BC}} + A_{V,\mathrm{ISM}}}$), and grayer attenuation curves ($\Delta\delta_{\mathrm{ISM}} = +0.08$), arising from preferential sampling of compact star-forming regions. Integrated fits underestimate stellar ages by $\sim23\%$ 
at $z < 5$ and 31$\%$ at $z \gtrsim 5$. The stellar mass offset correlates strongly with the age difference ($r = 0.78$) and the attenuation slope difference ($r = 0.84$), indicating that age-dependent outshining and spatially varying dust geometry are primary drivers of the discrepancy between resolved and integrated stellar masses.}
 {}


\maketitle
%

\section{Introduction}
Since its launch and first light earlier this decade, the James Webb Space Telescope (JWST) has been transforming our understanding of early galaxy evolution by enabling observations of rest-frame optical light at high redshifts \citep[e.g.,][]{Finkelstein2022, Castellano2022, Naidu2022,yan2023,Adams2023,javier23, Finkelstein23, gomezguijarro2023, Bunker2023, Sabti2024, Navarro-Carrera2024, Martorano2025, pg2025}. These observations have uncovered the underlying stellar mass and older stellar populations that remained largely obscured in earlier data from the Hubble Space Telescope (HST).

The synergy between JWST and HST now enables a more complete view of galaxies in the early Universe. HST's rest-frame ultraviolet (UV) imaging captures unattenuated star formation in galaxies beyond the cosmic noon, while the optical and near-infrared (NIR) capabilities of JWST probe obscured star formation and the evolved stellar component. At $z > 3$, the combination of UV and optical/NIR coverage is essential for breaking degeneracies between stellar age and dust attenuation. \citep[e.g.,][]{Wang2024, Iani2024, Weibel2024, Li2024}.

Together, the high spatial resolution of JWST and HST have opened a new window into the internal structure of high-redshift galaxies. HST provided some of the first resolved views of star-forming regions in the rest-frame UV \citep{Abraham99, LanyonFoster2012, Wuyts2013, Morishita2015, Abdurrouf2018, Jafariyazani2019, Lee2022}, while JWST extends this capability into the rest-frame optical and NIR, enabling spatially resolved mapping of stellar populations and star formation activity \citep{Polletta2024, Matharu2024}. Crucially, JWST allows for pixel-by-pixel (resolved) spectral energy distribution (SED) fitting across a broad wavelength range, capturing internal gradients and revealing sub-galactic variations in physical properties \citep{Wang2022, Song2023, Arteaga2023, pg2023, dEugenio2024}.\\

While resolved SED fitting with JWST represents a major leap forward in mapping internal physical properties of high-redshift galaxies, mounting evidence indicates that resolved and integrated SED fitting often yield systematically different stellar mass estimates \citep{Zibetti2009,Sorba2015,Sorba2018,pg2023,Arteaga2023, arteaga2024, Lines2025}. One primary mechanism driving these discrepancies is outshining \citep{Papovich2001,Conroy2013,Sorba2018, Tacchella2022, Topping2022, Suess2022, Whitler2023, Narayanan2024, Witten2025}. This effect arises when the luminous emission from young, massive stars dominates the integrated light, effectively masking the fainter contribution from older stellar populations, and thereby hindering accurate stellar mass estimates. Numerous studies have compared stellar masses derived from resolved and integrated photometry. While some report good agreement between the two approaches \citep[e.g.,][]{Hemmati2014,Cibinel2015,pg2023, Shen2024,Li2024,Lines2025}, others highlight that integrated SED fitting can systematically underestimate stellar mass due to outshining \citep[e.g.,][]{Pforr2012,Sorba2018,Arteaga2023}. Resolved SED fitting mitigates this bias by spatially disentangling star-forming regions from older stellar populations, which enables a more accurate reconstruction of the total stellar mass.\\

Assumptions about the star formation history (SFH) in SED fitting have been shown to significantly influence the severity of outshining-induced biases in stellar mass estimates \citep{Gallazzi2009, Bolzonella2010, Wuyts2013, Sorba2015, Narayanan2024, Jain2024}. Simplified parametric forms, such as exponentially declining SFHs, tend to overemphasize recent star formation activity, which in turn leads to systematic underestimation of both stellar masses and mass-weighted ages \citep{Narayanan2024, Jain2024}. Similarly, \citet{arteaga2024} found that single-component SFH models can underestimate the total stellar mass by up to 0.5 dex, whereas adopting a two-component SFH leads to better agreement with resolved measurements. \citet{Narayanan2024} showed that rigid SFH models fail to recover early star formation episodes when the observed light is dominated by recent bursts, and found that these biases can be mitigated by adopting flexible SFHs, that better reflect the diversity of star formation histories seen in high redshift galaxies \citep{Looser2025, Lisiecki25}.\\

While the influence of SFH on outshining and stellar mass biases has been widely explored \citep[e.g.,][]{Argumanez2023}, the role of dust attenuation in this context remains less understood. Dust attenuation occurs when interstellar dust grains absorb a substantial fraction of the UV and optical photons, predominantly emitted by young, massive stars, and re-emit this energy thermally in the infrared (IR), effectively redistributing the stellar radiation across the electromagnetic spectrum. The dust attenuation law has been shown to vary between galaxies (i.e., is non-universal) \citep{Kriek2013,Battisti2016, LoFaro2017, Salim2018, Malek2018, buat2019, Hamed23a, Markov2025, Shivaei2025}. The effect of dust attenuation on the SED of a galaxy, and therefore on the inferred physical properties, is influenced by various physical factors, such as dust composition \citep{Mascia2021}, metallicity \citep{Shivaei2020b,Shivaei2020a, Hamed23b}, and the relative distribution of stars and dust \citep{buat2019,Hamed23a}, all of which can shape the attenuation curve \citep[][]{Calzetti2000, CharlotFall2000, Narayanan2018, Salim2020}.\\

Given that both dust attenuation and SFH influence the relative visibility of young and old stellar populations, their combined effect may amplify or mitigate the outshining bias. This motivates a systematic investigation of how variations in dust attenuation properties affect the discrepancies between resolved and integrated stellar mass estimates, particularly in galaxies observed with JWST at high redshift.\\

Although previous spatially resolved studies have provided important insights, many focused on individual systems \citep[e.g.,][]{Arteaga2023}, limited to narrow redshift ranges \citep[e.g.,][]{Wang2022}, or biased toward the most massive galaxies \citep[e.g.,][]{pg2023}.
In this paper, we aim to analyze a statistically significant, mass-complete sample of galaxies across a wide redshift range ($3 \leq z < 9$), enabling a robust assessment of stellar mass estimation through resolved and integrated SED fitting, and the role dust attenuation plays in shaping these measurements. Using spatially resolved SED modeling based on JWST/NIRCam and HST/ACS imaging in the GOODS-South field, we specifically investigate how variations in the attenuation slope, and dust attenuation properties, correlate with discrepancies between resolved and integrated stellar mass estimates.\\

This paper is structured as follows. In Section \ref{sample}, we describe the data used in this work, as well as the sample selection criteria. In Section \ref{processing}, we describe the methods used in handling our data and processing the maps, such as matching the point spread functions (PSF) of the different bands in order to perform the resolved analysis. The techniques employed for the integrated and resolved SED fitting are presented in Section \ref{SED}. The results are presented in Section \ref{results}, and they are discussed and summarized in Section \ref{summary}.\\

Throughout this paper, we adopt a flat $\Lambda$CDM cosmology with $\Omega_\mathrm{M}$ = 0.3, $\Omega_\Lambda$ = 0.7, and a Hubble constant of H$_0$ = 70 km s$^{-1}$ Mpc$^{-1}$. Stellar masses and SED modeling are computed assuming a \citet{Chabrier2003} initial mass function (IMF).

\section{Dataset and sample selection}\label{sample}

\subsection{Sample description}

The sample used in this paper has been assembled from the \texttt{Astrodeep}-JWST catalogue \citep{Merlin2024}, in the Great Observatories Origins Deep Survey \citep[GOODS,][]{Giavalisco2004} South field \citep{Beckwith2006}. This catalogue provides source detection and photometry on JWST NIRCam imaging mosaics from the JWST Advanced Deep Extragalactic Survey \citep[JADES,][]{Eisenstein2023,Rieke2023, Eisenstein2025} Data Releases 1 and 2. The JADES imaging provides exceptional depth, reaching a 5$\sigma$ point-source detection limit of $m_{\mathrm{AB}} \approx 30$ mag in the F200W filter in the Deep regions \citep{Rieke2023, Eisenstein2025}. The GOODS-South field was selected due to its extensive multi-wavelength coverage and the availability of deep archival data, which is ideal for resolved studies of galaxies across cosmic time. Since our primary goal is to investigate how dust attenuation affects the discrepancies between resolved and integrated properties, particularly stellar mass estimates, we leverage the complementary capabilities of JWST and HST in this well-characterized field. This dataset allows us to probe both the rest-frame UV and optical emission of galaxies at high redshift.\\

The NIRCam mosaics for the GOODS-South region are drawn from the JADES program's Guaranteed Time Observation (GTO), specifically programs 1180 (PI: Eisenstein) and 1210 (PI: Luetzgendorf). In \texttt{Astrodeep}-JWST, NIRCam observations were supplemented with imaging from the Hubble Legacy Fields \citep[HLF,][]{Illingworth2016}.
This combines deep ACS/WFC imaging with NIRCam wide and medium bands for a total of eighteen broadbands (fourteen NIRCam bands and four ACS/WFC) spanning from $0.4$ to $5\ \mu$m \citep{Illingworth2016, Oesch2023, Williams2023}. All the maps were calibrated and drizzled to match the same angular scale of 0.03$\arcsec$ per pixel. The \texttt{Astrodeep}-JWST catalog for the GOODS-South field contains 73\,638 galaxies, with photometric redshifts computed as the mean of the two central values from four estimates: one from \texttt{ZPHOT} \citep{Fontana2000} and three from \texttt{EAzY} \citep{Brammer2008}, each using a different template configuration, to reach a reliable redshift estimation \citep{Merlin2024}.\\

In the \texttt{Astrodeep}-JWST catalogue, \citet{Merlin2024} measured photometry on PSF-matched images using fixed circular apertures, from which an optimal aperture is selected per source based on its segmentation area. This ensured robust total flux estimates for each galaxy in the parent sample.\\

To enable a comprehensive study of resolved physical properties, with a particular focus on dust attenuation effects, we adopted an inclusive sample selection strategy. Rather than applying strict cuts for representativeness, we prioritized ensuring sufficient photometric coverage for reliable SED fitting. Specifically, we required each galaxy to have detections in at least four photometric bands with a minimum signal-to-noise ratio (S/N) of 3$\sigma$ in each. 
This criterion resulted in 54\,125 galaxies selected from the parent sample. To ensure adequate sampling of the rest-frame UV and optical spectrum which is crucial for reliably constraining stellar ages, dust attenuation, and stellar masses \citep{Walcher2011}, we adopted a minimum redshift cut of $z = 3$. At these redshifts, the combined wavelength coverage of HST/ACS (0.4-0.8~$\mu$m) and JWST/NIRCam (0.9-5~$\mu$m) maps onto the rest-frame UV–optical window. This enables robust recovery of stellar population parameters, including the UV slope, the stellar ages, and mass. Applying this cut yields a sample of 13\,094 galaxies. We do not impose a strict upper redshift cut beyond our S/N requirements in order to explore the feasibility of spatially resolved SED analysis toward the earliest cosmic epochs. Our full sample reaches a maximum redshift of z = 8.8.\\

We then selected a stellar mass-complete subsample, by following the approach outlined by \citet{Pozzetti2010}, which estimates the limiting stellar mass for each galaxy based on its observed magnitude. Specifically, for each source, we calculate a limiting mass defined as the stellar mass a galaxy would have if its apparent magnitude were equal to the limiting magnitude of the survey in the F444W band. For each redshift bin, we identify the stellar mass above which 90$\%$ of galaxies exceed M$_{\mathrm{lim}}$ and adopt this as the 90$\%$ stellar mass completeness limit. This defines a mass-complete subsample for subsequent analysis while minimizing selection biases. The stellar masses for mass-completeness selection were obtained from \citet{Merlin2024}. Applying these criteria results in a sample of 3\,415 galaxies, complete down to $\log (\mathrm{M}_\star / \mathrm{M}_\odot) = 7.5$.\\

\begin{figure*}[t]
\centering
\includegraphics[trim = 0mm 0mm 0mm 0mm, width=1\textwidth]{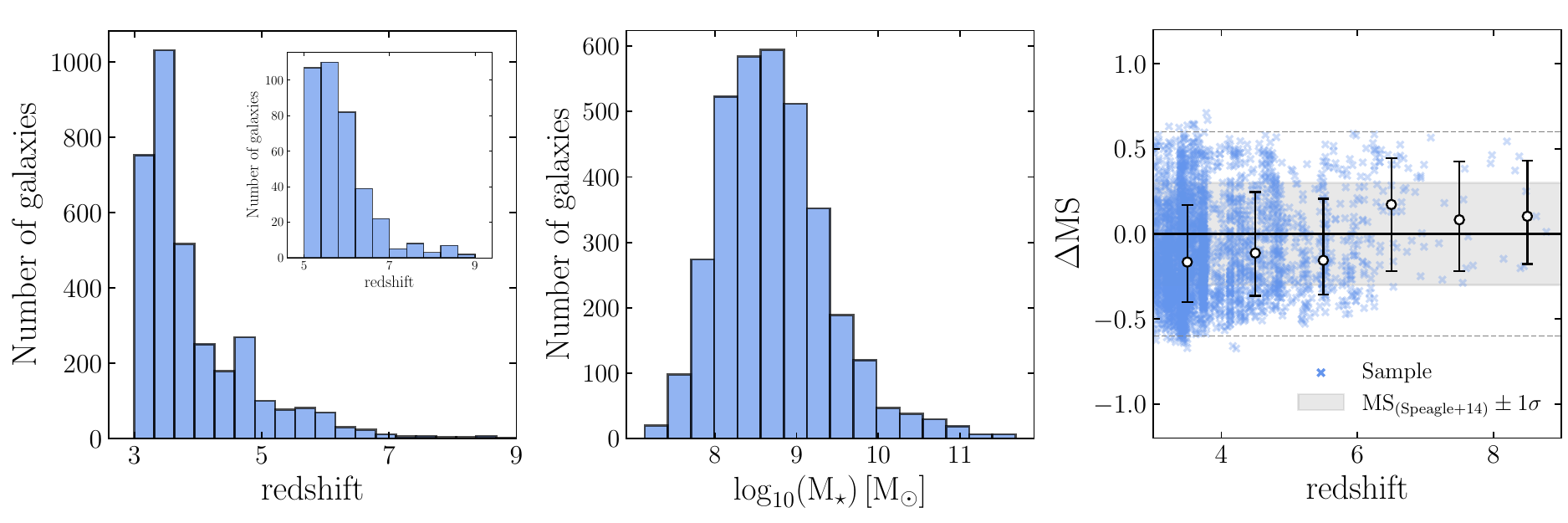}
\caption{Characterization of the galaxy sample used in this work. \textbf{Left}: redshift distribution of the sample. The distribution at $5 \leq z < 9$ is magnified in the inset for better visualization. \textbf{Middle}: distribution of stellar masses derived from SED fitting based on aperture-photometry fluxes for the sample from \citet{Merlin2024}. \textbf{Right}: offset from the star-forming main sequence, defined as $\Delta \mathrm{MS} = \log(\mathrm{SFR_{SED}}) - \log(\mathrm{SFR_{MS}}(M_{\star}, z))$, as a function of redshift, computed relative to the evolving main-sequence relation of \citet{Speagle2014}. Black circles indicate binned medians with 16th–84th percentile ranges. The shaded region denotes the $\pm1\sigma$ intrinsic scatter of the main sequence, and the dashed lines mark the $\pm0.6$~dex thresholds commonly used to separate starburst and quiescent regimes.}
\label{Fig.sample}
\end{figure*}

To minimize contamination from active galactic nuclei (AGN) emission, we cross-matched our sample with the Chandra Deep Field South X-ray catalog \citep{Luo2017}, identifying 19 X-ray sources. Of these, we excluded 7 bright AGN (intrinsic X-ray luminosity $L_X > 10^{44}$ erg s$^{-1}$) to prevent contamination of the stellar population fits, while retaining 12 sources with lower X-ray luminosities where AGN emission is expected to be negligible in the observed optical/NIR bands. We acknowledge that X-ray selection may not capture all AGN at high redshift \citep{Maiolino2025}. Our X-ray based AGN exclusion therefore represents a conservative approach that removes only the most luminous AGN, while lower-luminosity or X-ray weak AGN may remain in the sample. However, given that our sample is mass-complete and dominated by star-forming galaxies (98\% within the main sequence), any residual AGN contamination is expected to be minimal. This resulted in a final sample of 3\,408 galaxies, which form the basis for our study of both resolved and integrated SED properties. Figure \ref{Fig.sample} presents the main characteristics of the sample, including its redshift distribution, stellar masses, and offset from the star forming main sequence. The vast majority of galaxies (98$\%$) lie within the main sequence defined by \citet{Speagle2014}.

For the purposes of our analysis, we divide the sample into two redshift bins: $3 \leq z < 5$ and $5 \leq z$. This division reflects the evolving rest-frame wavelength coverage of the photometry with redshift. At $z > 5$, the reddest available band no longer probes the rest-frame NIR, which limits our sensitivity to the oldest stellar populations and can affect the robustness of stellar mass estimates. Our full sample reaches a maximum redshift of $z = 8.8$, with a median redshift of $z = 3.6$. The redshift interquartile range (middle 50$\%$ of the distribution) is $3.4 < z < 4.3$, and 90$\%$ of galaxies lie below $z = 5.2$. In total, $383$ galaxies lie in the $5 \leq z < 9$ bin (with interquartile in $5.3 < z < 6.2$), while the remaining galaxies fall within $3 \leq z < 5$ (with interquartile being $3.3 < z < 3.9$).

\subsection{Map processing}\label{processing}
To ensure consistent spatial resolution across all photometric bands prior to the resolved SED fitting, we performed PSF matching to the broadest NIRCam filter (F444W), which has the largest full width at half maximum (FWHM) measured in our images (0.16$\arcsec$). Empirical PSFs were constructed for each band using \texttt{Photutils} \citep{Bradley2024} by identifying and stacking isolated, unsaturated stars in the field. On average, 47 stars were selected independently per band to account for wavelength-dependent PSF variations, proper motions between epochs, and filter-specific detection limits. Each star was centered, normalized, and the final PSF was built as the weighted mean of the stack. To suppress noise from PSF wings and focus on the core structure, a circular mask was applied during stacking \citep[as done in][]{pg2023}.\\

We convolved each image to the F444W resolution using a custom-built kernel derived in Fourier space from the empirical PSFs. This method accelerates computation and maintains kernel accuracy. The choice of a tapering function is crucial in building the kernels, as it controls how the transformation is applied in Fourier space \citep{Martinache2020}.
We adopted a split cosine bell tapering function \citep{ Berkheimer2024,Matsuura2024}, which preserved the extended PSF structures. The split cosine bell function has two tapering parameters: $\alpha$ (defining the width of the flat central region) and $\beta$ (controlling the width of the transition region). 

Unlike a sharp frequency cutoff, which can introduce ringing artifacts, the split cosine bell provided a smooth transition to zero, preserving the core structure of the PSF while suppressing noisy components in the wings. This is especially important when matching heterogeneous datasets, such as HST to JWST or between NIRCam filters, where differences in spatial resolution and detector response can lead to significant variation in PSF structure.
We found that varying $\alpha$ between 0.4 and 0.6, and $\beta$ between 0.1 and 0.3 produced optimal results. The final FWHM of the convolved images were checked using the same stars, matching the FWHM measured for F444W.\\

We then measured photometry on a pixel-by-pixel basis using the PSF-matched JWST and HST maps (with a pixel scale corresponding to 0.03 arcsec per pixel), while measuring their noise empirically in source free regions of the original cutouts, to avoid the noise correlation introduced by the convolution. While correlated noise is still present at some level in the drizzled mosaics due to resampling, this effect is not explicitly corrected for. For the resolved SED fitting, we retained only those pixels that had at least a S/N of three in at least four photometric bands \citep{pg2023}. This mild criterion ensures that the resolved SED fitting is performed reliably only on pixels with sufficient photometric coverage. This pixel selection resulted in 138 pixels per galaxy on average. Given the matched PSF FWHM of 0.16" and pixel scale of 0.03"/pixel, this corresponds to approximately 6 independent spatial resolution elements per galaxy. To ensure a fair comparison with integrated measurements, we quantified the area covered by the pixels used in the resolved analysis relative to the total aperture adopted for each galaxy. On average, the selected pixels cover 35$\%$ of the total aperture area measured by \citet{Merlin2024}, with an interquartile range of 29-40$\%$, and extending up to 46$\%$ at the 90th percentile across the sample. These pixels correspond to the brightest, high-S/N regions where the spatially resolved SED fits are most reliable. Despite representing only about one-third of the total aperture area, these regions recover a large fraction of the total flux: the ratio between the recovered and integrated fluxes has a median of 0.68, with an interquartile range of 0.59–0.76. The stellar masses derived from the resolved and integrated analysis (Section~\ref{SED}) were scaled by the corresponding aperture correction factors, which have a median value of 1.47 (with an interquartile range of 1.30–1.68) to account for the fraction of the total aperture not covered by the pixels used in the resolved analysis. This ensures that the total stellar masses reflect the full galaxy flux.

\section{Spectral energy distribution modeling}\label{SED}

\subsection{Spatially resolved SED fitting}
Traditionally, SED fitting has been performed on the integrated light of galaxies, treating them as spatially uniform systems. However, this approach overlooks internal variations in stellar populations, star formation, and dust content. Spatially resolved SED fitting, conducted on a pixel-by-pixel basis, provides a more detailed view of the underlying physical processes by capturing sub-galactic gradients in age, mass, and attenuation \citep{Argumanez2023,Arteaga2023, pg2023}. This method has become increasingly important for characterizing complex structures within galaxies.

To derive the spatially resolved physical properties of galaxies, we used the Code Investigating GALaxy Emission \citep[{\ttfamily CIGALE, version 2025},][]{Boquien2019}, which implements a Bayesian approach to SED fitting and allows flexible modeling of both star formation histories and attenuation laws. In this work, we configured {\ttfamily CIGALE} to model the stellar and dust emission of each spatial element using a range of physically motivated templates. The code allows for a modular construction of the SED, which enables the combination of different SFHs, attenuation laws, and nebular emission components. Below, we describe the specific modules and parameter choices adopted for our analysis.\\

\subsubsection{Stellar component}
To model the stellar emission, we adopted the \citet{BC03} single stellar population models, allowing metallicities to range from $0.01 \times \mathrm{Z}_\odot$ up to $\mathrm{Z_\odot}$. This was done to capture the internal diversity of galaxy regions (e.g., localized star-forming regions with sub-solar metallicities, and evolved regions with near-solar ones). Metallicity might influence the shape of the stellar continuum, particularly in the optical–NIR, therefore restricting metallicity to a narrow range (e.g., fixed solar) can lead to biased estimates of physical parameters \citep{Maraston2010}. Additionally, we accounted for nebular continuum and line emission in our SED fitting, since it is crucial in UV–optical bands, especially for young star-forming regions and high redshift galaxies \citep{Boquien2010, deBarros2014}.\\

A fundamental component of SED fitting is the SFH, as it governs the relative contribution of stellar populations of different ages and directly influences derived physical properties such as stellar mass, age, and star formation rate \citep{Wuyts2012, Conroy2013, Ciesla2016}. For this work, where our primary objective is to understand how dust attenuation influences stellar mass estimates in resolved fits, we performed a systematic exploration of several commonly adopted parametric SFH forms, including constant, delayed exponentially declining (SFR(t) $\propto$ t $\times$ exp(-t/$\tau$)), and delayed with a recent burst within the last 100 Myr.\\

We assessed the performance of each SFH model by comparing the resolved and integrated stellar masses, as well as the corresponding SEDs and reduced $\chi^2$ values. We find that the adopted SFH exerts an important influence on the derived stellar masses, particularly at lower stellar masses where recent star formation dominates the integrated light. Models assuming shorter minimum stellar ages or bursty histories tend to yield slightly larger mass discrepancies, consistent with the stronger impact of outshining by young stellar populations. Although the overall systematic offset between resolved and integrated estimates persists across all SFH assumptions, the amplitude and scatter of $\Delta log M_\star$ vary with the chosen SFH, underscoring the importance of the star-formation history in shaping the inferred stellar masses.\\

As described in Appendix~\ref{app_SFH}, we tested several SFH parametrizations including constant, delayed exponentially declining, and delayed with a recent burst, as well as different minimum stellar ages in the SSP models (10, 25, 50, and 100~Myr). This minimum stellar ages in the SSP models refer to the youngest simple stellar population age available in the model grid. The choice of SFH introduces systematic offsets in the resolved-to-integrated stellar mass difference $\Delta \log \mathrm{M}_\star$: median values range from 0.16~dex (delayed + burst) to 0.30~dex (constant SFH with 10~Myr minimum age), with a maximum variation of 0.14~dex across all tested configurations. We adopted a constant SFH with a minimum stellar age of 50~Myr, which yields an intermediate systematic offset of 0.24~dex and the lowest scatter ($\sigma = 0.18$~dex) among all tested models. This choice balances computational efficiency while avoiding unrealistically young stellar populations that could enhance outshining bias, and preserves robust stellar mass estimates. The constant SFH, despite its simplicity, has been shown to yield reliable stellar masses in spatially resolved analyses \citep[e.g.,][]{Arteaga2023, Lines2025}.

\subsubsection{Dust component}
Interstellar dust affects the studied wavelength regime in this work via the dust attenuation. Modeling dust attenuation is a crucial component of SED fitting, particularly for high-redshift galaxies, where dusty star-forming regions dominate UV and optical light. Despite major advancements in attenuation modeling, key challenges remain, especially in disentangling age and dust effects, understanding the geometry of stars and dust, and constraining the properties of the oldest stellar populations.\\

Dust attenuation in galaxies is often described using either empirical prescriptions or more physically motivated models. The widely used \citet{Calzetti2000} law treats attenuation as arising from a uniform foreground dust screen, providing a good empirical fit to the integrated light of typical star-forming galaxies \citep[e.g.,][]{Buat12,Malek2017,Elbaz2018, Ciesla2020, Hamed21}. However, this formalism lacks the flexibility to capture the greyer attenuation curves expected in systems with complex star-dust geometries, where age-dependent obscuration and spatial mixing become important \citep[e.g.,][]{Noll2009, buat2019, Salim2020}.\\

To better account for these complexities, we adopt the two-component dust attenuation model introduced by \citet{CharlotFall2000}. In this framework, stars younger than a characteristic timescale, set to 10 Myr, corresponding to the dispersal time of their natal birth clouds (BCs), are attenuated by both the dense BCs and the diffuse interstellar medium (ISM), while older stars experience attenuation from the ISM only. This age-dependent treatment reflects the physical distinction between newly formed stars embedded in their birth environments and the older stellar populations dispersed throughout the galaxy.

The attenuation at each wavelength is modeled using power-law functions for each component:

\begin{equation}\label{equation_av_ISM}
A_\lambda(t) = A_V^{\mathrm{ISM}} \left( \frac{\lambda}{5500\,\text{\AA}} \right)^{\delta_{\mathrm{ISM}}} + 
\begin{cases}
A_V^{\mathrm{BC}} \left( \frac{\lambda}{5500\,\text{\AA}} \right)^{\delta_{\mathrm{BC}}}, & \text{if } t < 10\ \mathrm{Myr} \\
0, & \text{if } t \geq 10\ \mathrm{Myr}
\end{cases}
\end{equation}

Here, $A_V^{\mathrm{ISM}}$ and $A_V^{\mathrm{BC}}$ are the $V$-band attenuations in the ISM and BC components, respectively, while $\delta_{\mathrm{ISM}}$ and $\delta_{\mathrm{BC}}$ control the steepness of the attenuation curve for each component. The relative contribution of the ISM and BC is parameterized by $\mu = A_V^{\mathrm{ISM}} / (A_V^{\mathrm{ISM}} + A_V^{\mathrm{BC}})$. The variable $t$ denotes the stellar age, and 10 Myr is the characteristic timescale for birth cloud dispersal.\\

We fix the slope of the birth cloud component to $\delta_{\mathrm{BC}} = -1.3$, a steep value consistent with the highly embedded environments around young stars \citep[as in e.g.,][]{Wild2007, daCunha2008, Chevallard2013}. In contrast, the ISM slope $\delta_{\mathrm{ISM}}$ is left as a free parameter to capture spatial variations in dust properties. The original \citet{CharlotFall2000} model adopted a fixed $\delta_{\mathrm{ISM}} = -0.7$ to reproduce the average attenuation curve derived by \citet{Calzetti2000}, but this fixed slope does not reflect the diversity of attenuation curves observed in galaxies. In particular, grayer attenuation curves that are linked to mixed geometries of stars and dust, have been reported in both local and high-redshift dusty galaxies \citep[e.g.,][]{Pierini2004, Chevallard2013, Salim2018, Trayford2020}. Allowing $\delta_{\mathrm{ISM}}$ to vary enables us to explore such deviations and assess how the wavelength dependence of attenuation varies across different regions within galaxies.\\

\begin{table}[]
\caption{List of input parameters for SED modeling with {\ttfamily CIGALE}. The parameter values shown here correspond to those of the final computation of the SEDs, after testing different value ranges. The values are linearly equally spaced.}\label{tab:sedparam}

  \begin{center}
 
    \begin{tabular}{l r}
     \hline\hline
    \textbf{Parameter} & \textbf{Priors}\\
     \hline\hline
     \multicolumn{2}{c}{Constant SFH}   \\
     \hline
    Age$_{\star}$ [Myr]   & 100 values in [50-2000]\\
    \hline\hline
     \multicolumn{2}{c}{Nebular emission \citep{Inoue2011}}   \\
     \hline
     Ionization parameter (log $U$)& -3, -2, -1\\
    \hline\hline
    \multicolumn{2}{c}{{SSP} \citep{BC03}}   \\
    \hline
    IMF & \citep{Chabrier2003}\\
    $\mathrm{Z_\star}$ &  3 values in [$Z_{\odot}$/100 , $Z_{\odot}$]\\
    Separation age$^\mathrm{i}$ &     10\,Myr\\
    \hline\hline
    \multicolumn{2}{c}{{Dust attenuation} \citep[modified][]{CharlotFall2000}}   \\
    \hline
    A(V)$_{\mathrm{ISM}}$ & 100 values in [0-3] \\
    $\mu^{\mathrm{ii}}$ &   7 values in [0.3-1]\\
    $\delta_{\mathrm{ISM}}$ & 7 values in [-1.5, -0.3]\\
    \hline
    \hline
    \end{tabular}
    \end{center}
    $^\mathrm{i}$ The separation age defines the threshold below which young stars are attenuated by both BC and the ISM, while older stars are only affected by ISM attenuation.\\
    $^{\mathrm{ii}}$ $A_V^{\mathrm{ISM}} / \left(A_V^{\mathrm{BC}} + A_V^{\mathrm{ISM}}\right)$.
    \end{table}
    
\subsection{Integrated SED fitting}
To enable a direct comparison with our spatially resolved analysis, we performed integrated SED fitting by summing the photometric fluxes from the same set of PSF-matched pixels used in the resolved analysis. Specifically, we included only those pixels that met the predefined S/N threshold of $3\sigma$ in at least four bands. This approach mirrors the methodology used in previous resolved studies of high-redshift galaxies \citep[e.g.,][]{Arteaga2023}. The stellar masses inferred from this integrated approach were also multiplied by the aperture correction factors discussed in Section~\ref{processing}. This ensured consistency between the resolved and integrated measurements by accounting for the flux outside the selected high S/N pixels. The resulting integrated SEDs thus represent the total emission within the same apertures adopted for the resolved analysis, which enables a one-to-one comparison of stellar masses derived from both methods. Appendix~\ref{SED_comparison} illustrates these systematic differences for an example galaxy, showing the spatial gradients in mass, age, and attenuation that integrated fitting averages over.\\

The SED fitting was carried out using the same configuration as in the resolved case, with the same set of templates, parameter priors, and physical assumptions. We adopted the \citet{BC03} stellar population models. Stellar metallicity was allowed to vary between sub-solar and solar values, and nebular emission lines were included. Dust attenuation was modeled using the \citet{CharlotFall2000} double power law, with the attenuation slope in the diffuse ISM left as a free parameter to account for variations in dust geometry and composition. Table \ref{tab:sedparam} shows the parameters used in both resolved and integrated SED fitting. By fitting the integrated SEDs with the same assumptions and selection criteria as the resolved fits, we are able to isolate the effects of spatial resolution and examine biases in the derived properties. \\

In the case of resolved SED fitting, physical properties were derived for each individual pixel. For the stellar mass and SFR, the values from all pixels within each galaxy were summed to obtain the total galaxy-integrated values, with uncertainties propagated in quadrature. Resolved mass-weighted stellar ages were computed as the mass-weighted mean 
across all pixels, $\langle\mathrm{Age}\rangle_M = \sum_i (M_{\star,i} \times 
\mathrm{Age}_i) / \sum_i M_{\star,i}$, where $i$ indexes individual pixels. The resolved dust attenuation in the $FUV$ and $V$ bands were computed by summing the attenuated and unattenuated rest-frame luminosities from the best-fit SEDs at 150\,nm and 550\,nm across all the pixels within each galaxy. The total fluxes were then used to derive the band-specific attenuation as

\begin{equation}\label{equation_Aresolved}
A_\lambda^{\mathrm{resolved}} = -2.5 \log_{10} \left( \frac{\sum\limits_{i} F_{\lambda, i}^{\mathrm{att}}}{\sum\limits_{i} F_{\lambda, i}^{\mathrm{int}}} \right),
\end{equation}
where $F_{\lambda, i}^{\mathrm{att}}$ is the attenuated flux of pixel $i$ at wavelength $\lambda$, and $F_{\lambda, i}^{\mathrm{int}}$ is the intrinsic (unattenuated) flux of pixel $i$. The uncertainty on \( A_\lambda^{\mathrm{resolved}} \) was derived using standard error propagation, assuming independence between pixel uncertainties.\\

To estimate the resolved attenuation curve slope of the ISM, we computed the attenuation contributed by the ISM component using the attenuated and unattenuated luminosities across all pixels in a set of broad-band filters spanning the rest-frame UV to NIR (150 to 1600 nm) as defined in Equation \ref{equation_Aresolved}. This yielded a set of $A_\lambda^{\mathrm{ISM}}$ values as a function of wavelength. We then fitted these values using the functional form of the ISM attenuation law in Equation \ref{equation_av_ISM}, treating the slope $\delta_{\mathrm{ISM}}$ as a free parameter. This yields the ISM attenuation curve slope for each galaxy.\\

To estimate the resolved UV slope $\beta$ \citep{Calzetti1994}, we computed the total rest-frame attenuated luminosities within ten narrow spectral windows spanning the wavelength range from 125 nm to 250 nm. These windows were selected to avoid strong spectral features and represent the underlying UV continuum \citep{Finkelstein2012, Rogers2013}. For each galaxy, we summed the attenuated fluxes of all valid pixels within each of the ten Calzetti windows. The UV slope $\beta$ was then obtained by fitting a power-law $F_\lambda \propto \lambda^\beta$ to the total fluxes as a function of wavelength across the windows.

\begin{figure}[t]
\centering
\includegraphics[trim = 0mm 0mm 0mm 0mm, width=0.5\textwidth]{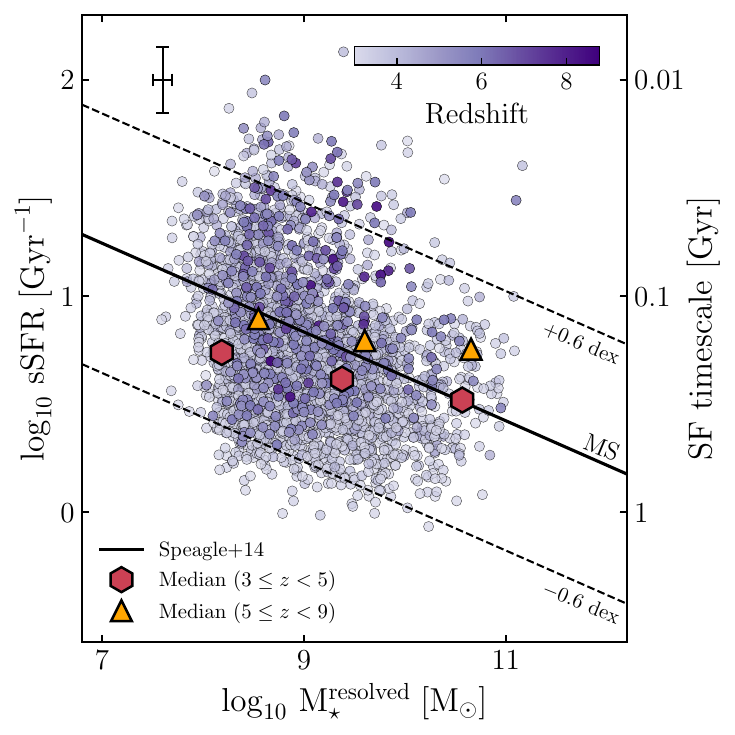}
\caption{Resolved specific star formation rate (sSFR) in Gyr$^{-1}$ as a function of resolved stellar mass for the galaxy sample used in this work, color-coded by redshift. Each point represents an individual galaxy. The $y$-axis on the right side shows the corresponding star-formation timescale ($1/\mathrm{sSFR}$) in Gyr. The solid black line indicates the star-forming main sequence (MS) from \citet{Speagle2014}, evaluated at $z \simeq 3.6$, while the dashed lines denote offsets of $\pm 0.6$ dex (a factor of four) relative to the MS. Hexagonal and triangular symbols mark the median sSFR in bins of stellar mass for galaxies in the redshift ranges $3 \leq z < 5$ and $5 \leq z < 9$, respectively. The representative median measurement uncertainty in $\log \mathrm{M}_\star$ and $\log \mathrm{sSFR}$ is shown by the black error bar in the upper-left corner.}
\label{Fig.sSFR}
\end{figure}

\subsection{Model quality assessment}
The quality of the SED fitting results was assessed using the reduced $\chi^2$ statistic. This metric quantifies the agreement between the model and the observed photometry while accounting for photometric uncertainties and the number of fitted parameters. We first explored parameter grids and inspected posterior probability density functions to identify degeneracies and constrain priors, with particular attention to age and dust attenuation parameters. The final SED calculation included priors that are listed in Table \ref{tab:sedparam}. To evaluate the consistency of the resolved approach, we summed the best-fit pixel-by-pixel model SEDs and compared the resulting total fluxes to the integrated observed photometry in each band. We find a median reduced $\chi^2$ of 1.79 for the resolved approach, compared to 0.80 for the integrated fits. The lower reduced $\chi^2$ values in the integrated approach reflect the fact that it directly optimizes the fit to the total galaxy photometry, whereas the resolved approach sums independently fitted pixels, which does not guarantee a globally optimal solution for the integrated light.\\
 
To further evaluate the robustness of the derived physical parameters, we conducted a mock analysis \citep{Osborne2024} and present its results in Appendix~\ref{app_mock}. This test assesses the ability of the adopted SED model configuration to recover input physical properties from synthetic photometry. The mock analysis confirms that key parameters such as stellar mass and SFR are reliably recovered, with small systematic offsets and tight correlations between the input and recovered values. These results validate the internal consistency of our SED modeling approach and highlight the strengths and limitations of parameter recovery under realistic observational conditions.\\

\section{Results $\&$ Discussion}\label{results}
Following the SED fitting in Section~\ref{SED}, resolved and integrated estimates were derived for stellar mass, SFR, mass-weighted age, and attenuation properties. Median values and dispersions in redshift bins are summarized in Table~\ref{tab:resolved_vs_integrated}.
\begin{figure}[]
\centering
\includegraphics[width=0.5\textwidth]{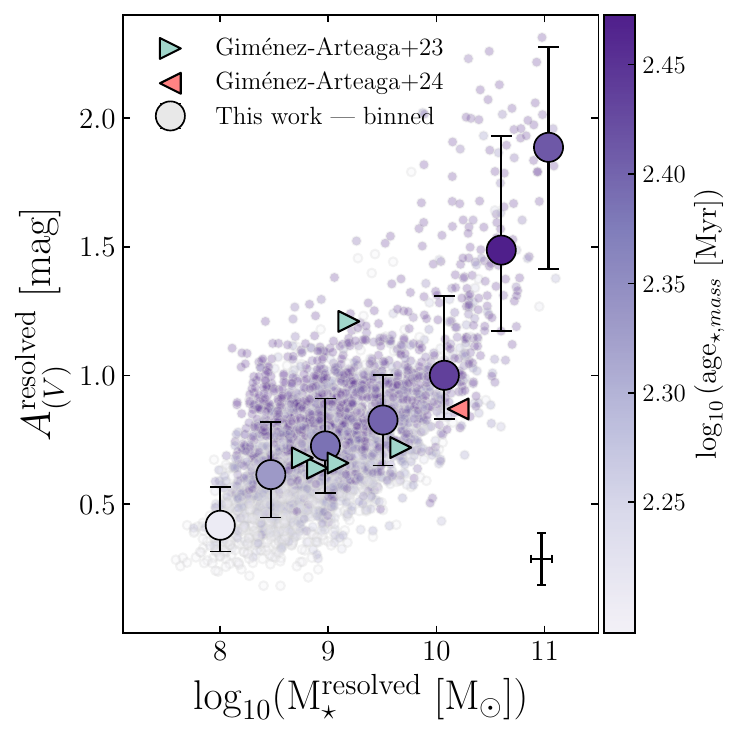}
\caption{Resolved dust attenuation $A(V)$ versus stellar mass, color-coded by mass-weighted stellar age. Large circles show binned medians with error bars indicating the 16th-84th percentile spread in $A(V)$. Literature comparisons from \citet{Arteaga2023} (turquoise triangles) and \citet{arteaga2024} (red triangle) are shown for reference. The error bar in the lower-right corner represents median measurement uncertainties.}
\label{Fig.AV_mass}
\end{figure}
Figure~\ref{Fig.sSFR} shows the relation between spatially resolved specific SFR (sSFR = SFR/M$_\star$) and stellar mass, colored by redshift, with the \citet{Speagle2014} main sequence used as the baseline. Here, SFR refers to the SED-derived instantaneous SFR, whereas the \citet{Speagle2014} relation is primarily calibrated using UV-based SFR indicators. The bulk of the sample lies within the main sequence band. Across the full sample, the median offset from the main sequence is $\Delta\log\mathrm{sSFR} = -0.16$ dex, indicating that galaxies in our sample have slightly lower sSFR than predicted by the main sequence relation, with an interquartile range of 0.40 dex. The majority of galaxies (94.8\%) fall within the main sequence boundaries defined by $-0.6 < \Delta \mathrm{log(sSFR)} < +0.6$ dex. Only a small fraction deviate significantly, with 2.7\% classified as starbursts ($\Delta \mathrm{log(sSFR)} \ge +0.6$ dex) and 2.5\% as quiescent systems ($\Delta \mathrm{log(sSFR)} \le -0.6$ dex).\\

Figure~\ref{Fig.AV_mass} shows the relation between spatially resolved $V$-band attenuation ($A_V$) and stellar mass, color-coded by the mass-weighted stellar age. The attenuation increases steadily with stellar mass, rising from $A_V\approx0.4$~mag in the lowest-mass bins to $\approx1.9$~mag at the high-mass end. Across the full sample, the median attenuation is $\tilde{A_V}=0.73$~mag, with an interquartile range of $0.58$-$0.89$~mag. Galaxies below $\log \mathrm{M}_\star/\mathrm{M_{\odot}}=9.5$ show a median $A_V=0.67$~mag, while those with $\log \mathrm{M}_\star/\mathrm{M_{\odot}}>9.5$ have a median $A_V=0.96$~mag. The observed $\sim$0.3 mag scatter around the $A_V$–$M_\star$ relation might reflect variations in dust geometry and stellar population mixing rather than observational uncertainties. The overall trend of increasing attenuation with stellar mass is consistent with the consensus found in previous high-redshift studies. At $z\simeq2$–3, \citet{McLure2018} showed that attenuation scales with stellar mass following a Calzetti-like law, with galaxies of $\log \mathrm{M}_\star/\mathrm{M_{\odot}}\gtrsim9.5$ typically reaching $A_V\sim1$ mag. Recent JWST-based analyses (\citealt{Markov2025}) extend this trend up to $z\sim12$. The resolved $A_V$–$M_\star$ relation in Figure~\ref{Fig.AV_mass} closely follows the one-dimensional relation derived from the aperture-photometry SED fits. At the high-mass end, the higher attenuations partly reflect that the resolved $A_V$ values are based on S/N-selected pixels, primarily sampling the bright central regions of galaxies, whereas the low-surface-brightness outskirts with lower attenuation are less represented in the measured flux-weighted attenuation. These results are also in agreement with \citet{Arteaga2023} and \citet{arteaga2024}, where they analyzed singular galaxies at $5 \leq z <9$ star-forming galaxies.\\

\begin{table*}
\centering
\caption{Median physical properties of the galaxy sample in different redshift intervals, comparing the results from the spatially resolved and integrated SED fitting analyses. Reported values correspond to the median and the 16th–84th percentile range of the distribution (1$\sigma$ scatter). Stellar masses and SFRs have been corrected for aperture effects as described in Section~\ref{processing}. Ages refer to mass-weighted stellar population ages in Myr, and $A_V$ represents the total $V$-band attenuation in magnitudes. 
The parameters $\mu$ and $\delta_{\mathrm{ISM}}$ describe, respectively, the fraction of total attenuation arising in the diffuse ISM and the slope of the ISM attenuation law.}
\renewcommand{\arraystretch}{1.15}
\resizebox{\textwidth}{!}{
\begin{tabular}{cccccccccc}
\hline\hline
\begin{tabular}[c]{@{}c@{}}redshift\end{tabular} &
\begin{tabular}[c]{@{}c@{}}$N$\end{tabular} &
 &
\begin{tabular}[c]{@{}c@{}}$\langle \log M_* \rangle$\\ {[}$M_\odot${]}\end{tabular} &
\begin{tabular}[c]{@{}c@{}}$\langle \mathrm{log\,SFR} \rangle$\\ {[}$M_\odot\,\mathrm{yr}^{-1}${]}\end{tabular} &
\begin{tabular}[c]{@{}c@{}}$\langle \mathrm{Age_{(mass)}} \rangle$\\ {[}Myr{]}\end{tabular} &
\begin{tabular}[c]{@{}c@{}}$\langle A_V \rangle$\\ {[}mag{]}\end{tabular} &
\begin{tabular}[c]{@{}c@{}}$\langle \mu \rangle$\\ \end{tabular} &
\begin{tabular}[c]{@{}c@{}}$\langle \delta_{\mathrm{ISM}} \rangle$\\ \end{tabular} \\
\hline
\rule{0pt}{2.6ex}
\multirow{2}{*}{$3 \le z < 4$} & \multirow{2}{*}{2327} & \rule{0pt}{3ex} Resolved   & $8.96^{+0.77}_{-0.53}$ & $0.7^{+0.7}_{-0.5}$ & $268.2^{+46.5}_{-73.7}$ & $0.92^{+0.11}_{-0.11}$ & $0.44 ^{+0.01}_{-0.01}$ & $-0.54 ^{+0.04}_{-0.04}$ \\
                               &                        & \rule{0pt}{3ex} Integrated & $8.75^{+0.71}_{-0.57}$ & $0.6^{+0.7}_{-0.5}$ & $209.4^{+108.8}_{-141.9}$ & $0.74^{+0.58}_{-0.58}$ & $ 0.72 ^{+0.08}_{-0.06}$ & $ -0.61 ^{+0.14}_{-0.24} $ \\[3pt]
\hline
\rule{0pt}{2.6ex}
\multirow{2}{*}{$4 \le z < 5$} & \multirow{2}{*}{696} & \rule{0pt}{3ex} Resolved   & $8.93^{+0.67}_{-0.45}$ & $0.9^{+0.6}_{-0.4}$ & $183.3^{+48.7}_{-78.9}$ & $0.88^{+0.12}_{-0.12}$ & $0.45 ^{+0.02}_{-0.01}$ & $-0.50 ^{+0.04}_{-0.05}$ \\
                               &                       & \rule{0pt}{3ex} Integrated & $8.72^{+0.63}_{-0.56}$ & $0.8^{+0.6}_{-0.5}$ & $130.5^{+103.9}_{-102.9}$ & $0.64^{+0.55}_{-0.55}$ & $0.74 ^{+0.10}_{-0.09}$ & $-0.61 ^{+0.13}_{-0.21}$ \\[3pt]
\hline
\rule{0pt}{2.6ex}
\multirow{2}{*}{$5 \le z < 6$} & \multirow{2}{*}{272} & \rule{0pt}{3ex} Resolved   & $8.98^{+0.66}_{-0.51}$ & $1.0^{+0.6}_{-0.5}$ & $159.5^{+23.1}_{-57.2}$ & $0.91^{+0.12}_{-0.12}$ & $ 0.45 ^{+0.01}_{-0.01}$ & $-0.46 ^{+0.02}_{-0.05}$ \\
                               &                       & \rule{0pt}{3ex} Integrated & $8.73^{+0.74}_{-0.60}$ & $0.9^{+0.8}_{-0.6}$ & $115.3^{+81.7}_{-91.0}$ & $0.89^{+0.66}_{-0.66}$ & $0.75 ^{+0.13}_{-0.08}$ & $-0.55 ^{+0.16}_{-0.20} $ \\[3pt]
\hline
\rule{0pt}{2.6ex}
\multirow{2}{*}{$6 \le z < 7$} & \multirow{2}{*}{88} & \rule{0pt}{3ex} Resolved   & $8.88^{+0.50}_{-0.37}$ & $1.0^{+0.5}_{-0.3}$ & $111.9^{+31.1}_{-37.4}$ & $0.89^{+0.13}_{-0.13}$ & $0.46 ^{+0.01}_{-0.01}$ & $-0.46 ^{+0.04}_{-0.03}$ \\
                               &                      & \rule{0pt}{3ex} Integrated & $8.60^{+0.61}_{-0.53}$ & $1.0^{+0.5}_{-0.5}$ & $55.6^{+91.2}_{-38.3}$ & $0.80^{+0.65}_{-0.65}$ & $0.81 ^{+0.09}_{-0.06}$ & $-0.52 ^{+0.09}_{-0.15} $ \\[3pt]
\hline
\rule{0pt}{2.6ex}
\multirow{2}{*}{$7 \le z < 8$} & \multirow{2}{*}{15} & \rule{0pt}{3ex} Resolved   & $9.37^{+0.25}_{-0.54}$ & $1.5^{+0.2}_{-0.6}$ & $110.7^{+3.9}_{-33.7}$ & $1.21^{+0.15}_{-0.15}$ & $0.46 ^{+0.01}_{-0.01}$ & $-0.47 ^{+0.02}_{-0.03}$ \\
                               &                      & \rule{0pt}{3ex} Integrated & $9.29^{+0.24}_{-0.72}$ & $1.5^{+0.4}_{-0.8}$ & $90.2^{+28.1}_{-69.1}$ & $1.40^{+0.73}_{-0.73}$ & $0.81 ^{+0.08}_{-0.13}$ & $-0.40 ^{+0.07}_{-0.10}$ \\[3pt]
\hline
\rule{0pt}{2.6ex}
\multirow{2}{*}{$8 \le z < 9$} & \multirow{2}{*}{10} & \rule{0pt}{3ex} Resolved   & $9.26^{+0.52}_{-0.26}$ & $1.4^{+0.5}_{-0.3}$ & $89.7^{+7.0}_{-7.7}$ & $1.09^{+0.14}_{-0.14}$ & $0.46 ^{+0.00}_{-0.01}$ & $-0.46 ^{+0.04}_{-0.01} $ \\
                               &                      & \rule{0pt}{3ex} Integrated & $9.10^{+0.49}_{-0.58}$ & $1.3^{+0.7}_{-0.6}$ & $70.4^{+26.1}_{-28.3}$ & $1.18^{+0.73}_{-0.73}$ & $0.82 ^{+0.06}_{-0.14}$ & $-0.47 ^{+0.07}_{-0.19}$ \\[3pt]
\hline
\end{tabular}}
\label{tab:resolved_vs_integrated}
\end{table*}

\subsection{Resolved versus integrated stellar masses}

We summarize in Table \ref{tab:resolved_vs_integrated} the median physical properties of our galaxy sample obtained from the spatially resolved and integrated SED fits across six redshift bins.

Figure~\ref{Fig.mass_comparison} compares the stellar masses derived from the resolved and integrated SED fits, color-coded by resolved dust attenuation properties. The resolved fits systematically yield higher stellar masses, but the offset exhibits a strong mass dependence. At low masses ($7 < \log \mathrm{M}_* [\mathrm{M}_{\odot}] < 8$), the offset is $\Delta\log \mathrm{M}_* = +0.34$ dex. The offset decreases with increasing mass: $+0.22$ dex at $\log \mathrm{M}_* [\mathrm{M}_{\odot}] \sim 8.5$, $+0.21$ dex at $\log \mathrm{M}_* [\mathrm{M}_{\odot}] \sim 9.5$, and approaching the 1:1 relation ($+0.04$ dex) at the highest masses ($\log \mathrm{M}_* [\mathrm{M}_{\odot}] \sim 10.5$). This mass-dependent behavior is consistent at both low and high redshifts. The global median offset of $\Delta\log \mathrm{M}_* = +0.24^{+0.16}_{-0.17}$ dex is dominated by intermediate-mass galaxies ($\log \mathrm{M}_* \sim 8$--10), which comprise 85\% of the sample.

This mass-dependent trend demonstrates that outshining disproportionately affects low-mass galaxies. In these systems, a few bright young stellar populations can dominate the integrated light, causing integrated SED fits to underestimate the total stellar mass by factors of $\sim$2 (0.3 dex at low masses). At higher masses ($\log M_* > 10$), older stellar populations contribute more significantly to the integrated light, reducing the bias and bringing resolved and integrated estimates into agreement. This stellar mass offset is consistent with the differences reported for massive systems in \citet{pg2023} and at $z\sim4$--6 in \citet{Lines2025}, where resolved and integrated stellar masses of normal star-forming galaxies agree within $\lesssim$ 0.3 dex.

We note that the systematic underestimation of stellar masses by integrated SED fitting, particularly at $\log M_\star/M_\odot < 9$ where we find offsets of $\sim$0.3 dex, may be relevant when interpreting black hole-to-stellar mass ratios at high redshift. Given that our X-ray-based AGN exclusion does not capture X-ray weak AGN \citep{Maiolino2025}, some fraction of our sample may host lower-luminosity AGN. In such cases, resolved stellar mass estimates would provide more accurate measurements for assessing black hole scaling relations \citep{Pacucci2023}.
\subsection{Dust attenuation and stellar population properties}

The stellar mass offset is accompanied by systematic differences in dust attenuation properties and stellar ages (Table~\ref{tab:resolved_vs_integrated}, 
Table~\ref{tab:delta_properties}). Resolved fits yield systematically higher $A_V$ values with a median offset of $\Delta A_V \approx +0.08$~mag up to $z\sim7$, lower $\mu$ ($\Delta\mu \approx -0.28$), and systematically older mass-weighted stellar ages. Relative to the resolved ages, the integrated fits underestimate stellar ages by approximately 25$\%$ at $z < 5$ and by 30$\%$ at $z \gtrsim 5$, corresponding to median absolute differences of $\Delta \mathrm{Age_{mass}} \approx 56$~Myr at lower redshifts and $\approx 35$~Myr at higher redshifts.

The higher $A_V$ values in resolved fits occur because they preferentially weight high surface brightness regions where young, attenuated stellar populations dominate the flux. These compact star-forming regions experience strong attenuation from both birth clouds and the diffuse ISM. In contrast, integrated SEDs average the light from both heavily attenuated central regions and less obscured outer regions, resulting in a lower effective $A_V$. The lower $\mu$ values \citep[$\mu \equiv \frac{A_{V,\mathrm{ISM}}}{A_{V,\mathrm{BC}} + A_{V,\mathrm{ISM}}}$,][]{CharlotFall2000} in the resolved fits further indicate that a larger fraction of the total attenuation arises in compact star-forming regions rather than in the diffuse ISM. In these regions, the optical depths are high \citep{Calzetti2000, Conroy2013, Mitchell2013}, which increases the intrinsic mass-to-light ratios locally. These systematic differences in $\mu$ between resolved ($\mu \sim 0.45$) and integrated ($\mu \sim 0.72$) fits suggest that when fitting integrated SEDs without spatial information, adopting lower $\mu$ priors (around 0.4--0.5) may better account for the contribution of compact birth clouds that are spatially resolved in pixel-by-pixel analysis but averaged over in integrated fits. This could help mitigate the systematic underestimation of stellar masses in unresolved observations. The resolved attenuation slopes ($\delta_{\mathrm{ISM}}$) are slightly shallower than in the integrated SEDs ($\Delta \delta_{\rm ISM} \approx 0.08$), reflecting that resolved fits preferentially sample compact, dusty regions where young stars are embedded in birth clouds (as indicated by lower $\mu$ values). In such clumpy, mixed star-dust geometries, the effective attenuation curve becomes grayer due to increased scattering and radiative transfer effects \citep{Trayford2020, Qin2024}. In contrast, integrated SEDs are averaged over regions with varying dust columns, producing steeper apparent slopes that dilute the contribution from birth clouds.

\begin{figure*}[]
\centering
\includegraphics[width=1\textwidth]{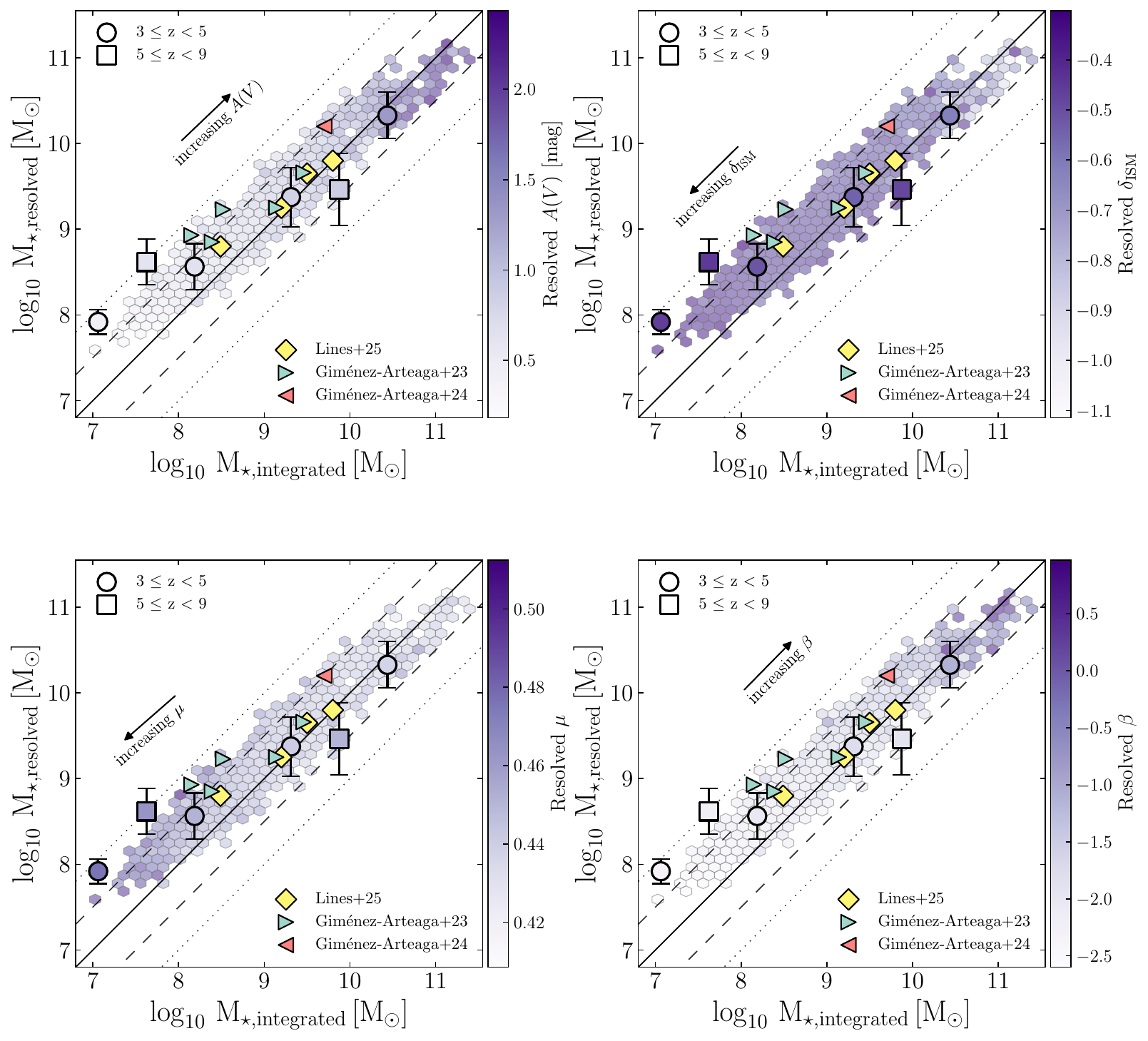}
\caption{Comparison between resolved and integrated stellar masses, corrected for aperture as described in Section~\ref{processing}, color-coded by the main attenuation parameters derived from the resolved SED fits: total $A_V$ (top left), ISM attenuation slope $\delta_{\mathrm{ISM}}$ (top right), $\mu$ (bottom left; the fraction of total attenuation arising in the diffuse ISM), and the UV continuum slope $\beta$ (bottom right). 
Each panel shows the relation between $\log M_{\star,\mathrm{resolved}}$ and $\log M_{\star,\mathrm{integrated}}$ for galaxies at $3 \leq z < 5$ (circles) and $5 \leq z < 9$ (squares), with median binned values shown in color and error bars indicating the $1\sigma$ dispersion. Solid, dashed, and dotted lines mark the one-to-one relation and offsets of $\pm 0.5$ and $\pm 1.0$ dex, respectively. The arrow indicates the direction of increasing attenuation parameter. Yellow squares, turquoise triangles, and red triangles mark, respectively, the measurements from \citet{Lines2025}, \citet{Arteaga2023}, and \citet{arteaga2024}.}
\label{Fig.mass_comparison}
\end{figure*}

These results illustrate the combined effects of outshining \citep{Papovich2001, Conroy2013, Narayanan2024} and spatially varying dust attenuation on integrated SED estimates. When using integrated photometry to analyze the stellar content of galaxies, the total light is dominated by the bright, young, and less obscured stellar populations, which outshine the older and dustier components. This bias leads the integrated SED fits to infer systematically younger stellar ages, lower $A_V$, and smaller intrinsic mass-to-light ratios, resulting in an underestimation of the total stellar mass. The resolved analysis mitigates this effect \citep{Sorba2015, Harvey2025} by fitting each pixel independently, thus recovering the contribution from older, more attenuated stellar populations located in different regions of the galaxy. The resolved $A_V$--$M_\star$ relation in Figure \ref{Fig.AV_mass} supports this interpretation, showing that within galaxies, regions of higher stellar mass exhibit stronger attenuation, consistent with dust being more concentrated in massive, central regions \citep{Wuyts2012, Tacchella2015}, which are downweighted in the integrated fits. By performing SED fitting on spatially resolved scales, the analysis spatially disentangles heavily attenuated star-forming regions from older, less obscured stellar populations. This spatial separation breaks the age-dust degeneracy that affects integrated fits, where emission from young, bright stars in compact birth clouds can dominate the integrated light and mask the contribution from older stellar populations distributed more diffusely across galaxies.

The correlations between $\Delta\log M_\star \equiv \log\!\left(\frac{M_{\star,\mathrm{resolved}}}{M_{\star,\mathrm{integrated}}}\right)$ and the other parameters provide quantitative support for these trends (Figure \ref{Fig.mass_comparison}). Across redshift bins, $\Delta\log M_\star$ shows the strongest linear correlation with the difference in mass-weighted stellar age (Pearson correlation coefficient $r=0.78$, $p=0.07$). A similarly strong correlation is found with the ISM attenuation slope difference ($r=0.84$, $p=0.04$), indicating that integrated fits assuming steeper attenuation laws reproduce the observed colors with insufficient $A_V$ and overly young stellar ages, thereby yielding lower $M/L$ ratios and underestimated stellar masses, whereas the resolved fits recover shallower slopes, higher attenuations, and correspondingly larger total stellar masses. Correlations with $\Delta A_V$ ($r=0.68$, $p=0.14$) and $\Delta\mu$ ($r=0.36$, $p=0.49$) suggest that variations in total attenuation and in the fraction of diffuse ISM attenuation contribute to the stellar mass offsets, though their influence appears less important than that of the slope and stellar ages. The weak dependence on $\Delta\log\mathrm{SFR}$ ($r=0.20$, $p=0.70$) supports the interpretation that the discrepancies between resolved and integrated stellar masses are driven primarily by the combined effects of age-dependent outshining and spatially inhomogeneous dust attenuation.\\

At higher redshifts ($z \gtrsim 5$), the age offset increases to 30\%, though the absolute difference decreases to $\Delta\mathrm{Age_{mass}} \approx 35$ Myr. This reflects both the younger overall stellar populations at these epochs (median $\sim$110 Myr vs. $\sim$265 Myr at z < 4) and the increasing photometric limitations as NIRCam filters no longer sample rest-frame NIR wavelengths. The mass offset remains significant at $\sim0.25$--0.28~dex. These trends likely reflect the increasing limitations of the observed photometry \citep{Whitler2023}, as the rest-frame NIR is no longer sampled by NIRCam filters. The systematically younger stellar ages at $z > 5$ (median $\sim$110 Myr vs. $\sim$265 Myr at $z < 4$) reduce the contribution from older stellar populations, potentially mitigating outshining effects even as photometric constraints weaken. Without these longer-wavelength constraints, SED fits become dominated by rest-frame UV light from young stellar populations, enhancing outshining effects and reducing sensitivity to the older, more massive components. Consequently, both resolved and integrated fits converge toward younger ages and lower inferred $A_V$, with integrated fits remaining more affected by this bias.

Positive $\Delta A_V$ and $\Delta\mathrm{Age_{mass}}$ values in Table~\ref{tab:resolved_vs_integrated}, together with the resolved $A_V$--$M_\star$ relation in Figure~\ref{Fig.AV_mass}, indicate that attenuation and stellar age both increase toward regions of higher stellar mass, which mostly corresponds to the central parts of galaxies, supporting the presence of strong internal $A_V$ and age gradients that are averaged out in integrated SED fits.

We find no significant dependence of the stellar mass offset on stellar metallicity, discussed in Section \ref{SED}. The correlation between $\Delta\log \mathrm{M}_\star$ and the stellar metallicity ($\mathrm{Z_\star}$) is weak ($r = 0.10$, $p < 10^{-3}$), indicating that variations in metallicity do not drive the differences between resolved and integrated stellar mass estimates. This is in line with works investigating stellar populations and SED \citep{Shapley2001, Conroy2013}.


\begin{table}
\centering
\caption{Median differences in attenuation properties (resolved $-$ integrated) in two resolved stellar mass bins.}
\renewcommand{\arraystretch}{1.15}
\small
\begin{tabular}{ccccc}
\hline\hline
Resolved $\log \mathrm{M}_*$ [M$_\odot$] & $N$ & $\Delta A_V$ [mag] & $\Delta \delta_{\rm ISM}$ & $\Delta \mu$ \\
\hline
$7.5 \le \log M_* < 9$   & 1846 & $+0.07$ & $+0.10$ & $-0.30$ \\
$9 \le \log M_* < 11.5$  & 1562 & $+0.08$ & $+0.08$ & $-0.26$ \\
\hline
\end{tabular}
\label{tab:delta_properties}
\end{table}

\section{Summary \& Conclusions}\label{summary}
In this work, we conducted spatially resolved SED analysis of 3\,408 mass-complete galaxies spanning $3 \le z < 9$ in GOODS-South, utilizing deep JWST/NIRCam and HST/ACS observations. Our analysis quantifies systematic biases between resolved and integrated stellar mass measurements and examines the role of dust geometry and stellar population properties in driving these discrepancies.\\

The resolved-to-integrated stellar mass comparison reveals a mass-dependent systematic offset. Low-mass systems ($\log M_\star/M_\odot < 8$) show the largest discrepancy at $\Delta\log M_\star = +0.34$ dex, which diminishes to $+0.22$ dex for $\log M_\star/M_\odot \sim 8.5$, then $+0.21$ dex at $\log M_\star/M_\odot \sim 9.5$, ultimately converging to near-unity ($+0.04$ dex) for the most massive galaxies ($\log M_\star/M_\odot \sim 10.5$). The sample-wide median offset of $\Delta\log M_\star = +0.24^{+0.16}_{-0.17}$ dex primarily reflects the intermediate-mass population ($\log M_\star \sim 8$--10) constituting 85\% of our sample. This systematic trend confirms that low-mass galaxies suffer most severely from outshining-induced biases, with integrated methods underestimating true masses by factors of $\sim$2 (0.3 dex). Conversely, massive systems ($\log M_\star > 10$) show excellent agreement between methods as older populations dominate their spectral signatures. These patterns arise from the interplay of outshining \citep{Sorba2018, Wuyts2013, Tacchella2022, Whitler2023} and heterogeneous dust distributions.

Systematic differences extend beyond mass to encompass dust and age parameters. Resolved measurements preferentially probe compact, dust-rich star-forming complexes \citep{Chevance2020}, yielding elevated attenuations (median $\Delta A_V \approx +0.08$~mag across mass bins) and reduced ISM dominance ($\Delta\mu$ from $-0.30$ to $-0.26$) relative to integrated estimates. The attenuation law slope exhibits graying ($\Delta\delta_{\mathrm{ISM}} = +0.08$), reflecting clumpy birth cloud geometries, while integrated approaches smooth over these structures producing artificially steep slopes. Age determinations show parallel offsets: resolved mass-weighted ages systematically exceed integrated values \citep{Wuyts2012}, with integrated methods underestimating by 25\% ($z < 5$) and 30\% ($z \gtrsim 5$), corresponding to absolute differences of $\approx 56$~Myr and $\approx 35$~Myr respectively. This age bias reveals integrated fitting's systematic preference for young, luminous components.

Correlation analysis identifies primary drivers of mass discrepancies. The age offset shows the strongest correlation with $\Delta\log M_\star$ ($r = 0.78$, $p = 0.07$), confirming age-dependent outshining as a dominant mechanism. The ISM slope difference exhibits comparable correlation strength ($r = 0.84$, $p = 0.04$), demonstrating that integrated fitting's preference for steep attenuation curves leads to insufficient $A_V$, spuriously young ages, reduced mass-to-light ratios, and consequently underestimated masses. The $\Delta A_V$ correlation ($r = 0.68$, $p = 0.14$) further implicates dust geometry in shaping inferred parameters. Resolved fitting's ability to spatially separate heavily attenuated young populations in birth clouds from diffuse older populations breaks the age-dust degeneracy, enabling recovery of obscured stellar components contributing negligibly to integrated spectra.

At extreme redshifts ($z \gtrsim 5$), offsets diminish from $\sim0.25$--0.28~dex ($z\lesssim6$) to $\lesssim0.1$ dex ($z>7$), likely reflecting photometric limitations as NIRCam filters no longer sample rest-frame NIR wavelengths. Without long-wavelength constraints, fitting becomes UV-dominated, amplifying outshining while reducing sensitivity to massive, evolved populations. Both methods consequently converge toward younger ages and lower $A_V$, though integrated approaches remain more severely biased.\\

Key limitations include S/N-based pixel selection necessarily excluding low-signal regions, which preferentially samples luminous central zones potentially under-representing extended low-surface-brightness populations. Our aperture corrections assume similar mass-to-light ratios between detected and excluded regions; systematic age or dust differences could introduce uncertainties. JWST/NIRSpec spatially resolved spectroscopy would provide independent age and dust constraints, directly validating SED-derived parameters.

Practical recommendations for integrated SED fitting emerge from our analysis: expect mass-dependent systematic underestimation ($\sim0.3$ dex for $\log M_\star < 8$, $\sim0.2$ dex at $\log M_\star \sim 8$--10, negligible beyond $\log M_\star > 10.5$). Integrated fitting preferentially adopts high $\mu$ values, over-emphasizing diffuse ISM attenuation. Lower $\mu$ priors (0.4--0.5) better capture birth cloud contributions, improving mass recovery. Flexible attenuation slopes enhance fitting quality, particularly for unresolved observations lacking direct dust geometry constraints.

Our findings demonstrate that systematic discrepancies between resolved and integrated stellar mass estimates fundamentally arise from the coupling of outshining with spatially varying dust attenuation geometry. Internal dust-light architecture must be considered when interpreting integrated SEDs, especially in low-mass, high-redshift regimes where biases are most pronounced. Resolved SED fitting, increasingly viable with JWST, offers a robust methodology for accurate mass determination while breaking age-dust degeneracies inherent to integrated approaches.

\begin{acknowledgements}
M.H. and P.G.P.-G. acknowledge support from grant PID2022-139567NB-I00 funded by Spanish Ministerio de Ciencia e Innovaci\'on MCIN/AEI/10.13039/501100011033, FEDER {\it Una manera de hacer Europa}. L.C. and S.A. acknowledge support from grant PID2021-127718NB-I00 funded by Spanish Ministerio de Ciencia e Innovaci\'on MCIN/AEI/10.13039/501100011033. A.J.B. and J.C. acknowledge funding from the ``FirstGalaxies'' Advanced Grant from the European Research Council (ERC) under the European Union's Horizon 2020 research and innovation program (Grant agreement No. 789056). C.N.A.W. and Z.J. acknowledge support from JWST/NIRCam contract to the University of Arizona NAS5-02105. H.\"U. acknowledges funding by the European Union (ERC APEX, 101164796). Views and opinions expressed are however those of the authors only and do not necessarily reflect those of the European Union or the European Research Council Executive Agency. Neither the European Union nor the granting authority can be held responsible for them. E.C.-L. acknowledges support of an STFC Webb Fellowship (ST/W001438/1). C.W. acknowledges support from NOIRLab, which is managed by the Association of Universities for Research in Astronomy (AURA) under a cooperative agreement with the National Science Foundation. M.H. thanks Tiago Cortinhal for the discussions about computational optimization. This work is based on observations made with the NASA/ESA/CSA James Webb Space Telescope. The data were obtained from the Mikulski Archive for Space Telescopes at the Space Telescope Science Institute, which is operated by the Association of Universities for Research in Astronomy, Inc., under NASA contract NAS 5-03127 for JWST. The authors acknowledge the teams of programs 1895 and 1963 for developing their observing program with a zero-exclusive-access period. This work has made use of the Rainbow Cosmological Surveys Database, which is operated by the Centro de Astrobiolog\'ia (CAB), CSIC-INTA.
\end{acknowledgements}

\bibliographystyle{aa}
\bibliography{aanda.bib}

@ARTICLE{Malek2018,
       author = {{Ma{\l}ek}, K. and {Buat}, V. and {Roehlly}, Y. and {Burgarella}, D. and
         {Hurley}, P.~D. and {Shirley}, R. and {Duncan}, K. and
         {Efstathiou}, A. and {Papadopoulos}, A. and {Vaccari}, M. and
         {Farrah}, D. and {Marchetti}, L. and {Oliver}, S.},
        title = "{HELP: modelling the spectral energy distributions of Herschel detected galaxies in the ELAIS N1 field}",
      journal = {\aap},
         year = 2018,
        month = nov,
       volume = {620},
          eid = {A50},
        pages = {A50},
          doi = {10.1051/0004-6361/201833131},
archivePrefix = {arXiv},
       eprint = {1809.00529},
 primaryClass = {astro-ph.GA},
       adsurl = {https://ui.adsabs.harvard.edu/abs/2018A&A...620A..50M}
}

@ARTICLE{Boquien2019,
       author = {{Boquien}, M. and {Burgarella}, D. and {Roehlly}, Y. and {Buat}, V. and
         {Ciesla}, L. and {Corre}, D. and {Inoue}, A.~K. and {Salas}, H.},
        title = "{CIGALE: a python Code Investigating GALaxy Emission}",
      journal = {\aap},
         year = 2019,
        month = feb,
       volume = {622},
          eid = {A103},
        pages = {A103},
          doi = {10.1051/0004-6361/201834156},
archivePrefix = {arXiv},
       eprint = {1811.03094},
 primaryClass = {astro-ph.GA},
       adsurl = {https://ui.adsabs.harvard.edu/abs/2019A&A...622A.103B}
}

@ARTICLE{BC03,
       author = {{Bruzual}, G. and {Charlot}, S.},
        title = "{Stellar population synthesis at the resolution of 2003}",
      journal = {\mnras},
         year = 2003,
        month = oct,
       volume = {344},
       number = {4},
        pages = {1000-1028},
          doi = {10.1046/j.1365-8711.2003.06897.x},
archivePrefix = {arXiv},
       eprint = {astro-ph/0309134},
 primaryClass = {astro-ph},
       adsurl = {https://ui.adsabs.harvard.edu/abs/2003MNRAS.344.1000B}
}

@ARTICLE{Chabrier2003,
       author = {{Chabrier}, Gilles},
        title = "{Galactic Stellar and Substellar Initial Mass Function}",
      journal = {\pasp},
         year = 2003,
        month = jul,
       volume = {115},
       number = {809},
        pages = {763-795},
          doi = {10.1086/376392},
archivePrefix = {arXiv},
       eprint = {astro-ph/0304382},
 primaryClass = {astro-ph},
       adsurl = {https://ui.adsabs.harvard.edu/abs/2003PASP..115..763C}
}

@ARTICLE{Ciesla2016,
       author = {{Ciesla}, L. and {Boselli}, A. and {Elbaz}, D. and {Boissier}, S. and
         {Buat}, V. and {Charmandaris}, V. and {Schreiber}, C. and
         {B{\'e}thermin}, M. and {Baes}, M. and {Boquien}, M. and
         {De Looze}, I. and {Fern{\'a}ndez-Ontiveros}, J.~A. and
         {Pappalardo}, C. and {Spinoglio}, L. and {Viaene}, S.},
        title = "{The imprint of rapid star formation quenching on the spectral energy distributions of galaxies}",
      journal = {\aap},
         year = 2016,
        month = jan,
       volume = {585},
          eid = {A43},
        pages = {A43},
          doi = {10.1051/0004-6361/201527107},
archivePrefix = {arXiv},
       eprint = {1510.07657},
 primaryClass = {astro-ph.GA},
       adsurl = {https://ui.adsabs.harvard.edu/abs/2016A&A...585A..43C}
}

@ARTICLE{Calzetti2000,
       author = {{Calzetti}, Daniela and {Armus}, Lee and {Bohlin}, Ralph C. and
         {Kinney}, Anne L. and {Koornneef}, Jan and {Storchi-Bergmann}, Thaisa},
        title = "{The Dust Content and Opacity of Actively Star-forming Galaxies}",
      journal = {\apj},
         year = 2000,
        month = apr,
       volume = {533},
       number = {2},
        pages = {682-695},
          doi = {10.1086/308692},
archivePrefix = {arXiv},
       eprint = {astro-ph/9911459},
 primaryClass = {astro-ph},
       adsurl = {https://ui.adsabs.harvard.edu/abs/2000ApJ...533..682C}
}

@ARTICLE{CharlotFall2000,
       author = {{Charlot}, St{\'e}phane and {Fall}, S. Michael},
        title = "{A Simple Model for the Absorption of Starlight by Dust in Galaxies}",
      journal = {\apj},
         year = 2000,
        month = aug,
       volume = {539},
       number = {2},
        pages = {718-731},
          doi = {10.1086/309250},
archivePrefix = {arXiv},
       eprint = {astro-ph/0003128},
 primaryClass = {astro-ph},
       adsurl = {https://ui.adsabs.harvard.edu/abs/2000ApJ...539..718C}
}

@ARTICLE{Ciesla2020,
       author = {{Ciesla}, L. and {B{\'e}thermin}, M. and {Daddi}, E. and {Richard}, J. and
         {Diaz-Santos}, T. and {Sargent}, M.~T. and {Elbaz}, D. and
         {Boquien}, M. and {Wang}, T. and {Schreiber}, C. and {Yang}, C. and
         {Zabl}, J. and {Fraser}, M. and {Aravena}, M. and {Assef}, R.~J. and
         {Baker}, A.~J. and {Beelen}, A. and {Boselli}, A. and {Bournaud}, F. and
         {Burgarella}, D. and {Charmandaris}, V. and {C{\^o}t{\'e}}, P. and
         {Epinat}, B. and {Ferrarese}, L. and {Gobat}, R. and {Ilbert}, O.},
        title = "{A hyper luminous starburst at z = 4.72 magnified by a lensing galaxy pair at z = 1.48}",
      journal = {\aap},
         year = 2020,
        month = mar,
       volume = {635},
          eid = {A27},
        pages = {A27},
          doi = {10.1051/0004-6361/201936727},
archivePrefix = {arXiv},
       eprint = {2001.03641},
 primaryClass = {astro-ph.GA},
       adsurl = {https://ui.adsabs.harvard.edu/abs/2020A&A...635A..27C}
}

@ARTICLE{Elbaz2018,
       author = {{Elbaz}, D. and {Leiton}, R. and {Nagar}, N. and {Okumura}, K. and
         {Franco}, M. and {Schreiber}, C. and {Pannella}, M. and {Wang}, T. and
         {Dickinson}, M. and {D{\'\i}az-Santos}, T. and {Ciesla}, L. and
         {Daddi}, E. and {Bournaud}, F. and {Magdis}, G. and {Zhou}, L. and
         {Rujopakarn}, W.},
        title = "{Starbursts in and out of the star-formation main sequence}",
      journal = {\aap},
         year = 2018,
        month = aug,
       volume = {616},
          eid = {A110},
        pages = {A110},
          doi = {10.1051/0004-6361/201732370},
archivePrefix = {arXiv},
       eprint = {1711.10047},
 primaryClass = {astro-ph.GA},
       adsurl = {https://ui.adsabs.harvard.edu/abs/2018A&A...616A.110E}
}

@ARTICLE{Malek2017,
       author = {{Ma{\l}ek}, K. and {Bankowicz}, M. and {Pollo}, A. and {Buat}, V. and
         {Takeuchi}, T.~T. and {Burgarella}, D. and {Goto}, T. and {Malkan}, M. and
         {Matsuhara}, H.},
        title = "{[Ultra] luminous infrared galaxies selected at 90 {\ensuremath{\mu}}m in the AKARI deep field: a study of AGN types contributing to their infrared emission}",
      journal = {\aap},
         year = 2017,
        month = feb,
       volume = {598},
          eid = {A1},
        pages = {A1},
          doi = {10.1051/0004-6361/201527969},
archivePrefix = {arXiv},
       eprint = {1611.07410},
 primaryClass = {astro-ph.GA},
       adsurl = {https://ui.adsabs.harvard.edu/abs/2017A&A...598A...1M}
}

@ARTICLE{LoFaro2017,
       author = {{Lo Faro}, B. and {Buat}, V. and {Roehlly}, Y. and
         {Alvarez-Marquez}, J. and {Burgarella}, D. and {Silva}, L. and
         {Efstathiou}, A.},
        title = "{Characterizing the UV-to-NIR shape of the dust attenuation curve of IR luminous galaxies up to z {\ensuremath{\sim}} 2}",
      journal = {\mnras},
         year = 2017,
        month = dec,
       volume = {472},
       number = {2},
        pages = {1372-1391},
          doi = {10.1093/mnras/stx1901},
archivePrefix = {arXiv},
       eprint = {1707.09805},
 primaryClass = {astro-ph.GA},
       adsurl = {https://ui.adsabs.harvard.edu/abs/2017MNRAS.472.1372L}
}

@ARTICLE{Salim2018,
       author = {{Salim}, Samir and {Boquien}, M{\'e}d{\'e}ric and {Lee}, Janice C.},
        title = "{Dust Attenuation Curves in the Local Universe: Demographics and New Laws for Star-forming Galaxies and High-redshift Analogs}",
      journal = {\apj},
         year = 2018,
        month = may,
       volume = {859},
       number = {1},
          eid = {11},
        pages = {11},
          doi = {10.3847/1538-4357/aabf3c},
archivePrefix = {arXiv},
       eprint = {1804.05850},
 primaryClass = {astro-ph.GA},
       adsurl = {https://ui.adsabs.harvard.edu/abs/2018ApJ...859...11S}
}

@ARTICLE{Buat12,
       author = {{Buat}, V. and {Noll}, S. and {Burgarella}, D. and {Giovannoli}, E. and
         {Charmandaris}, V. and {Pannella}, M. and {Hwang}, H.~S. and
         {Elbaz}, D. and {Dickinson}, M. and {Magdis}, G. and {Reddy}, N. and
         {Murphy}, E.~J.},
        title = "{GOODS-Herschel: dust attenuation properties of UV selected high redshift galaxies}",
      journal = {\aap},
         year = 2012,
        month = sep,
       volume = {545},
          eid = {A141},
        pages = {A141},
          doi = {10.1051/0004-6361/201219405},
archivePrefix = {arXiv},
       eprint = {1207.3528},
 primaryClass = {astro-ph.CO},
       adsurl = {https://ui.adsabs.harvard.edu/abs/2012A&A...545A.141B}
}

@ARTICLE{buat2019,
       author = {{Buat}, V. and {Ciesla}, L. and {Boquien}, M. and {Ma{\l}ek}, K. and
         {Burgarella}, D.},
        title = "{Cold dust and stellar emissions in dust-rich galaxies observed with ALMA: a challenge for SED-fitting techniques}",
      journal = {\aap},
         year = 2019,
        month = dec,
       volume = {632},
          eid = {A79},
        pages = {A79},
          doi = {10.1051/0004-6361/201936643},
       adsurl = {https://ui.adsabs.harvard.edu/abs/2019A&A...632A..79B}
}

@ARTICLE{Noll2009,
       author = {{Noll}, S. and {Burgarella}, D. and {Giovannoli}, E. and {Buat}, V. and
         {Marcillac}, D. and {Mu{\~n}oz-Mateos}, J.~C.},
        title = "{Analysis of galaxy spectral energy distributions from far-UV to far-IR with CIGALE: studying a SINGS test sample}",
      journal = {\aap},
         year = 2009,
        month = dec,
       volume = {507},
       number = {3},
        pages = {1793-1813},
          doi = {10.1051/0004-6361/200912497},
archivePrefix = {arXiv},
       eprint = {0909.5439},
 primaryClass = {astro-ph.CO},
       adsurl = {https://ui.adsabs.harvard.edu/abs/2009A&A...507.1793N}
}

@ARTICLE{daCunha2008,
       author = {{da Cunha}, Elisabete and {Charlot}, St{\'e}phane and {Elbaz}, David},
        title = "{A simple model to interpret the ultraviolet, optical and infrared emission from galaxies}",
      journal = {\mnras},
         year = 2008,
        month = aug,
       volume = {388},
       number = {4},
        pages = {1595-1617},
          doi = {10.1111/j.1365-2966.2008.13535.x},
archivePrefix = {arXiv},
       eprint = {0806.1020},
 primaryClass = {astro-ph},
       adsurl = {https://ui.adsabs.harvard.edu/abs/2008MNRAS.388.1595D}
}

@ARTICLE{Conroy2013,
       author = {{Conroy}, Charlie},
        title = "{Modeling the Panchromatic Spectral Energy Distributions of Galaxies}",
      journal = {\araa},
         year = 2013,
        month = aug,
       volume = {51},
       number = {1},
        pages = {393-455},
          doi = {10.1146/annurev-astro-082812-141017},
archivePrefix = {arXiv},
       eprint = {1301.7095},
 primaryClass = {astro-ph.CO},
       adsurl = {https://ui.adsabs.harvard.edu/abs/2013ARA&A..51..393C}
}

@ARTICLE{Salim2020,
       author = {{Salim}, Samir and {Narayanan}, Desika},
        title = "{The Dust Attenuation Law in Galaxies}",
      journal = {\araa},
         year = 2020,
        month = aug,
       volume = {58},
        pages = {529-575},
          doi = {10.1146/annurev-astro-032620-021933},
archivePrefix = {arXiv},
       eprint = {2001.03181},
 primaryClass = {astro-ph.GA},
       adsurl = {https://ui.adsabs.harvard.edu/abs/2020ARA&A..58..529S}
}

@ARTICLE{McLure2018,
       author = {{McLure}, R.~J. and {Dunlop}, J.~S. and {Cullen}, F. and {Bourne}, N. and {Best}, P.~N. and {Khochfar}, S. and {Bowler}, R.~A.~A. and {Biggs}, A.~D. and {Geach}, J.~E. and {Scott}, D. and {Micha{\l}owski}, M.~J. and {Rujopakarn}, W. and {van Kampen}, E. and {Kirkpatrick}, A. and {Pope}, A.},
        title = "{Dust attenuation in 2 < z < 3 star-forming galaxies from deep ALMA observations of the Hubble Ultra Deep Field}",
      journal = {\mnras},
         year = 2018,
        month = may,
       volume = {476},
       number = {3},
        pages = {3991-4006},
          doi = {10.1093/mnras/sty522},
archivePrefix = {arXiv},
       eprint = {1709.06102},
 primaryClass = {astro-ph.GA},
       adsurl = {https://ui.adsabs.harvard.edu/abs/2018MNRAS.476.3991M}
}

@ARTICLE{Kriek2013,
       author = {{Kriek}, Mariska and {Conroy}, Charlie},
        title = "{The Dust Attenuation Law in Distant Galaxies: Evidence for Variation with Spectral Type}",
      journal = {\apjl},
         year = 2013,
        month = sep,
       volume = {775},
       number = {1},
          eid = {L16},
        pages = {L16},
          doi = {10.1088/2041-8205/775/1/L16},
archivePrefix = {arXiv},
       eprint = {1308.1099},
 primaryClass = {astro-ph.CO},
       adsurl = {https://ui.adsabs.harvard.edu/abs/2013ApJ...775L..16K}
}

@ARTICLE{Shivaei2020b,
       author = {{Shivaei}, Irene and {Darvish}, Behnam and {Sattari}, Zahra and {Chartab}, Nima and {Mobasher}, Bahram and {Scoville}, Nick and {Rieke}, George},
        title = "{Dependence of the IRX-{\ensuremath{\beta}} Dust Attenuation Relation on Metallicity and Environment}",
      journal = {\apjl},
         year = 2020,
        month = nov,
       volume = {903},
       number = {2},
          eid = {L28},
        pages = {L28},
          doi = {10.3847/2041-8213/abc1ef},
archivePrefix = {arXiv},
       eprint = {2010.10538},
 primaryClass = {astro-ph.GA},
       adsurl = {https://ui.adsabs.harvard.edu/abs/2020ApJ...903L..28S}
}

@ARTICLE{Hamed21,
       author = {{Hamed}, M. and {Ciesla}, L. and {B{\'e}thermin}, M. and {Ma{\l}ek}, K. and {Daddi}, E. and {Sargent}, M.~T. and {Gobat}, R.},
        title = "{Multiwavelength dissection of a massive heavily dust-obscured galaxy and its blue companion at z{\ensuremath{\sim}}2}",
      journal = {\aap},
         year = 2021,
        month = feb,
       volume = {646},
          eid = {A127},
        pages = {A127},
          doi = {10.1051/0004-6361/202039577},
archivePrefix = {arXiv},
       eprint = {2101.07724},
 primaryClass = {astro-ph.GA},
       adsurl = {https://ui.adsabs.harvard.edu/abs/2021A&A...646A.127H}
}

@ARTICLE{Calzetti1994,
       author = {{Calzetti}, Daniela and {Kinney}, Anne L. and {Storchi-Bergmann}, Thaisa},
        title = "{Dust Extinction of the Stellar Continua in Starburst Galaxies: The Ultraviolet and Optical Extinction Law}",
      journal = {\apj},
         year = 1994,
        month = jul,
       volume = {429},
        pages = {582},
          doi = {10.1086/174346},
       adsurl = {https://ui.adsabs.harvard.edu/abs/1994ApJ...429..582C}
}

@ARTICLE{Hamed23b,
       author = {{Hamed}, M. and {Pistis}, F. and {Figueira}, M. and {Ma{\l}ek}, K. and {Nanni}, A. and {Buat}, V. and {Pollo}, A. and {Vergani}, D. and {Bolzonella}, M. and {Junais} and {Krywult}, J. and {Takeuchi}, T. and {Riccio}, G. and {Moutard}, T.},
        title = "{Decoding the IRX-{\ensuremath{\beta}} dust attenuation relation in star-forming galaxies at intermediate redshift}",
      journal = {\aap},
         year = 2023,
        month = nov,
       volume = {679},
          eid = {A26},
        pages = {A26},
          doi = {10.1051/0004-6361/202346976},
archivePrefix = {arXiv},
       eprint = {2309.01819},
 primaryClass = {astro-ph.GA},
       adsurl ={https://ui.adsabs.harvard.edu/abs/2023A&A...679A..26H}
}

@ARTICLE{Hamed23a,
       author = {{Hamed}, M. and {Ma{\l}ek}, K. and {Buat}, V. and {Junais} and {Ciesla}, L. and {Donevski}, D. and {Riccio}, G. and {Figueira}, M.},
        title = "{The slippery slope of dust attenuation curves. Correlation of dust attenuation laws with star-to-dust compactness up to z = 4}",
      journal = {\aap},
         year = 2023,
        month = jun,
       volume = {674},
          eid = {A99},
        pages = {A99},
          doi = {10.1051/0004-6361/202245818},
archivePrefix = {arXiv},
       eprint = {2304.13713},
 primaryClass = {astro-ph.GA},
       adsurl = {https://ui.adsabs.harvard.edu/abs/2023A&A...674A..99H}
}

@ARTICLE{Finkelstein2022,
       author = {{Finkelstein}, Steven L. and {Bagley}, Micaela B. and {Arrabal Haro}, Pablo and {Dickinson}, Mark and {Ferguson}, Henry C. and {Kartaltepe}, Jeyhan S. and {Papovich}, Casey and {Burgarella}, Denis and {Kocevski}, Dale D. and {Huertas-Company}, Marc and {Iyer}, Kartheik G. and {Koekemoer}, Anton M. and {Larson}, Rebecca L. and {P{\'e}rez-Gonz{\'a}lez}, Pablo G. and {Rose}, Caitlin and {Tacchella}, Sandro and {Wilkins}, Stephen M. and {Chworowsky}, Katherine and {Medrano}, Aubrey and {Morales}, Alexa M. and {Somerville}, Rachel S. and {Yung}, L.~Y. Aaron and {Fontana}, Adriano and {Giavalisco}, Mauro and {Grazian}, Andrea and {Grogin}, Norman A. and {Kewley}, Lisa J. and {Kirkpatrick}, Allison and {Kurczynski}, Peter and {Lotz}, Jennifer M. and {Pentericci}, Laura and {Pirzkal}, Nor and {Ravindranath}, Swara and {Ryan}, Russell E. and {Trump}, Jonathan R. and {Yang}, Guang and {Almaini}, Omar and {Amor{\'\i}n}, Ricardo O. and {Annunziatella}, Marianna and {Backhaus}, Bren E. and {Barro}, Guillermo and {Behroozi}, Peter and {Bell}, Eric F. and {Bhatawdekar}, Rachana and {Bisigello}, Laura and {Bromm}, Volker and {Buat}, V{\'e}ronique and {Buitrago}, Fernando and {Calabr{\`o}}, Antonello and {Casey}, Caitlin M. and {Castellano}, Marco and {Ch{\'a}vez Ortiz}, {\'O}scar A. and {Ciesla}, Laure and {Cleri}, Nikko J. and {Cohen}, Seth H. and {Cole}, Justin W. and {Cooke}, Kevin C. and {Cooper}, M.~C. and {Cooray}, Asantha R. and {Costantin}, Luca and {Cox}, Isabella G. and {Croton}, Darren and {Daddi}, Emanuele and {Dav{\'e}}, Romeel and {de La Vega}, Alexander and {Dekel}, Avishai and {Elbaz}, David and {Estrada-Carpenter}, Vicente and {Faber}, Sandra M. and {Fern{\'a}ndez}, Vital and {Finkelstein}, Keely D. and {Freundlich}, Jonathan and {Fujimoto}, Seiji and {Garc{\'\i}a-Argum{\'a}nez}, {\'A}ngela and {Gardner}, Jonathan P. and {Gawiser}, Eric and {G{\'o}mez-Guijarro}, Carlos and {Guo}, Yuchen and {Hamblin}, Kurt and {Hamilton}, Timothy S. and {Hathi}, Nimish P. and {Holwerda}, Benne W. and {Hirschmann}, Michaela and {Hutchison}, Taylor A. and {Jaskot}, Anne E. and {Jha}, Saurabh W. and {Jogee}, Shardha and {Juneau}, St{\'e}phanie and {Jung}, Intae and {Kassin}, Susan A. and {Le Bail}, Aur{\'e}lien and {Leung}, Gene C.~K. and {Lucas}, Ray A. and {Magnelli}, Benjamin and {Mantha}, Kameswara Bharadwaj and {Matharu}, Jasleen and {McGrath}, Elizabeth J. and {McIntosh}, Daniel H. and {Merlin}, Emiliano and {Mobasher}, Bahram and {Newman}, Jeffrey A. and {Nicholls}, David C. and {Pandya}, Viraj and {Rafelski}, Marc and {Ronayne}, Kaila and {Santini}, Paola and {Seill{\'e}}, Lise-Marie and {Shah}, Ekta A. and {Shen}, Lu and {Simons}, Raymond C. and {Snyder}, Gregory F. and {Stanway}, Elizabeth R. and {Straughn}, Amber N. and {Teplitz}, Harry I. and {Vanderhoof}, Brittany N. and {Vega-Ferrero}, Jes{\'u}s and {Wang}, Weichen and {Weiner}, Benjamin J. and {Willmer}, Christopher N.~A. and {Wuyts}, Stijn and {Zavala}, Jorge A. and {Ceers Team}},
        title = "{A Long Time Ago in a Galaxy Far, Far Away: A Candidate z = 12 Galaxy in Early JWST CEERS Imaging}",
      journal = {\apjl},
         year = 2022,
        month = dec,
       volume = {940},
       number = {2},
          eid = {L55},
        pages = {L55},
          doi = {10.3847/2041-8213/ac966e},
archivePrefix = {arXiv},
       eprint = {2207.12474},
 primaryClass = {astro-ph.GA},
       adsurl = {https://ui.adsabs.harvard.edu/abs/2022ApJ...940L..55F}
}

@ARTICLE{Rieke2023,
       author = {{Rieke}, Marcia J. and {Robertson}, Brant and {Tacchella}, Sandro and {Hainline}, Kevin and {Johnson}, Benjamin D. and {Hausen}, Ryan and {Ji}, Zhiyuan and {Willmer}, Christopher N.~A. and {Eisenstein}, Daniel J. and {Pusk{\'a}s}, D{\'a}vid and {Alberts}, Stacey and {Arribas}, Santiago and {Baker}, William M. and {Baum}, Stefi and {Bhatawdekar}, Rachana and {Bonaventura}, Nina and {Boyett}, Kristan and {Bunker}, Andrew J. and {Cameron}, Alex J. and {Carniani}, Stefano and {Charlot}, Stephane and {Chevallard}, Jacopo and {Chen}, Zuyi and {Curti}, Mirko and {Curtis-Lake}, Emma and {Danhaive}, A. Lola and {DeCoursey}, Christa and {Dressler}, Alan and {Egami}, Eiichi and {Endsley}, Ryan and {Helton}, Jakob M. and {Hviding}, Raphael E. and {Kumari}, Nimisha and {Looser}, Tobias J. and {Lyu}, Jianwei and {Maiolino}, Roberto and {Maseda}, Michael V. and {Nelson}, Erica J. and {Rieke}, George and {Rix}, Hans-Walter and {Sandles}, Lester and {Saxena}, Aayush and {Sharpe}, Katherine and {Shivaei}, Irene and {Skarbinski}, Maya and {Smit}, Renske and {Stark}, Daniel P. and {Stone}, Meredith and {Suess}, Katherine A. and {Sun}, Fengwu and {Topping}, Michael and {{\"U}bler}, Hannah and {Villanueva}, Natalia C. and {Wallace}, Imaan E.~B. and {Williams}, Christina C. and {Willott}, Chris and {Whitler}, Lily and {Witstok}, Joris and {Woodrum}, Charity},
        title = "{JADES Initial Data Release for the Hubble Ultra Deep Field: Revealing the Faint Infrared Sky with Deep JWST NIRCam Imaging}",
      journal = {\apjs},
         year = 2023,
        month = nov,
       volume = {269},
       number = {1},
          eid = {16},
        pages = {16},
          doi = {10.3847/1538-4365/acf44d},
archivePrefix = {arXiv},
       eprint = {2306.02466},
 primaryClass = {astro-ph.GA},
       adsurl = {https://ui.adsabs.harvard.edu/abs/2023ApJS..269...16R}
}

@ARTICLE{Sabti2024,
       author = {{Sabti}, Nashwan and {Mu{\~n}oz}, Julian B. and {Kamionkowski}, Marc},
        title = "{Insights from HST into Ultramassive Galaxies and Early-Universe Cosmology}",
      journal = {\prl},
         year = 2024,
        month = feb,
       volume = {132},
       number = {6},
          eid = {061002},
        pages = {061002},
          doi = {10.1103/PhysRevLett.132.061002},
archivePrefix = {arXiv},
       eprint = {2305.07049},
 primaryClass = {astro-ph.CO},
       adsurl = {https://ui.adsabs.harvard.edu/abs/2024PhRvL.132f1002S}
}

@ARTICLE{Jain2024,
       author = {{Jain}, Shweta and {Tacchella}, Sandro and {Mosleh}, Moein},
        title = "{The motivation for flexible star-formation histories from spatially resolved scales within galaxies}",
      journal = {\mnras},
         year = 2024,
        month = jan,
       volume = {527},
       number = {2},
        pages = {3291-3305},
          doi = {10.1093/mnras/stad3333},
archivePrefix = {arXiv},
       eprint = {2310.18462},
 primaryClass = {astro-ph.GA},
       adsurl = {https://ui.adsabs.harvard.edu/abs/2024MNRAS.527.3291J}
}

@ARTICLE{Suess2022,
       author = {{Suess}, Katherine A. and {Leja}, Joel and {Johnson}, Benjamin D. and {Bezanson}, Rachel and {Greene}, Jenny E. and {Kriek}, Mariska and {Lower}, Sidney and {Narayanan}, Desika and {Setton}, David J. and {Spilker}, Justin S.},
        title = "{Recovering the Star Formation Histories of Recently Quenched Galaxies: The Impact of Model and Prior Choices}",
      journal = {\apj},
         year = 2022,
        month = aug,
       volume = {935},
       number = {2},
          eid = {146},
        pages = {146},
          doi = {10.3847/1538-4357/ac82b0},
archivePrefix = {arXiv},
       eprint = {2207.02883},
 primaryClass = {astro-ph.GA},
       adsurl = {https://ui.adsabs.harvard.edu/abs/2022ApJ...935..146S}
}

@ARTICLE{Polletta2024,
       author = {{Polletta}, M. and {Frye}, B.~L. and {Garuda}, N. and {Willner}, S.~P. and {Berta}, S. and {Kneissl}, R. and {Dole}, H. and {Jansen}, R.~A. and {Lehnert}, M.~D. and {Cohen}, S.~H. and {Summers}, J. and {Windhorst}, R.~A. and {D'Silva}, J.~C.~J. and {Koekemoer}, A.~M. and {Coe}, D. and {Conselice}, C.~J. and {Driver}, S.~P. and {Grogin}, N.~A. and {Marshall}, M.~A. and {Nonino}, M. and {Ortiz}, III, R. and {Pirzkal}, N. and {Robotham}, A. and {Ryan}, R.~E. and {Willmer}, C.~N.~A. and {Yan}, H. and {Arumugam}, V. and {Cheng}, C. and {Gim}, H.~B. and {Hathi}, N.~P. and {Holwerda}, B. and {Kamieneski}, P. and {Keel}, W.~C. and {Li}, J. and {Pascale}, M. and {Rottgering}, H. and {Smith}, B.~M. and {Yun}, M.~S.},
        title = "{JWST's PEARLS: Resolved study of the stellar and dust components in starburst galaxies at cosmic noon}",
      journal = {\aap},
         year = 2024,
        month = oct,
       volume = {690},
          eid = {A285},
        pages = {A285},
          doi = {10.1051/0004-6361/202450671},
archivePrefix = {arXiv},
       eprint = {2405.07986},
 primaryClass = {astro-ph.GA},
       adsurl = {https://ui.adsabs.harvard.edu/abs/2024A&A...690A.285P}
}

@ARTICLE{Martinache2020,
       author = {{Martinache}, Frantz and {Ceau}, Alban and {Laugier}, Romain and {Kammerer}, Jens and {N'Diaye}, Mamadou and {Mary}, David and {Cvetojevic}, Nick and {Lopez}, Coline},
        title = "{Kernel-phase analysis: Aperture modeling prescriptions that minimize calibration errors}",
      journal = {\aap},
         year = 2020,
        month = apr,
       volume = {636},
          eid = {A72},
        pages = {A72},
          doi = {10.1051/0004-6361/201936981},
archivePrefix = {arXiv},
       eprint = {2003.02032},
 primaryClass = {astro-ph.IM},
       adsurl = {https://ui.adsabs.harvard.edu/abs/2020A&A...636A..72M}
}

@ARTICLE{Finkelstein23,
       author = {{Finkelstein}, Steven L. and {Bagley}, Micaela B. and {Ferguson}, Henry C. and {Wilkins}, Stephen M. and {Kartaltepe}, Jeyhan S. and {Papovich}, Casey and {Yung}, L.~Y. Aaron and {Arrabal Haro}, Pablo and {Behroozi}, Peter and {Dickinson}, Mark and {Kocevski}, Dale D. and {Koekemoer}, Anton M. and {Larson}, Rebecca L. and {Le Bail}, Aur{\'e}lien and {Morales}, Alexa M. and {P{\'e}rez-Gonz{\'a}lez}, Pablo G. and {Burgarella}, Denis and {Dav{\'e}}, Romeel and {Hirschmann}, Michaela and {Somerville}, Rachel S. and {Wuyts}, Stijn and {Bromm}, Volker and {Casey}, Caitlin M. and {Fontana}, Adriano and {Fujimoto}, Seiji and {Gardner}, Jonathan P. and {Giavalisco}, Mauro and {Grazian}, Andrea and {Grogin}, Norman A. and {Hathi}, Nimish P. and {Hutchison}, Taylor A. and {Jha}, Saurabh W. and {Jogee}, Shardha and {Kewley}, Lisa J. and {Kirkpatrick}, Allison and {Long}, Arianna S. and {Lotz}, Jennifer M. and {Pentericci}, Laura and {Pierel}, Justin D.~R. and {Pirzkal}, Nor and {Ravindranath}, Swara and {Ryan}, Russell E. and {Trump}, Jonathan R. and {Yang}, Guang and {Bhatawdekar}, Rachana and {Bisigello}, Laura and {Buat}, V{\'e}ronique and {Calabr{\`o}}, Antonello and {Castellano}, Marco and {Cleri}, Nikko J. and {Cooper}, M.~C. and {Croton}, Darren and {Daddi}, Emanuele and {Dekel}, Avishai and {Elbaz}, David and {Franco}, Maximilien and {Gawiser}, Eric and {Holwerda}, Benne W. and {Huertas-Company}, Marc and {Jaskot}, Anne E. and {Leung}, Gene C.~K. and {Lucas}, Ray A. and {Mobasher}, Bahram and {Pandya}, Viraj and {Tacchella}, Sandro and {Weiner}, Benjamin J. and {Zavala}, Jorge A.},
        title = "{CEERS Key Paper. I. An Early Look into the First 500 Myr of Galaxy Formation with JWST}",
      journal = {\apjl},
         year = 2023,
        month = mar,
       volume = {946},
       number = {1},
          eid = {L13},
        pages = {L13},
          doi = {10.3847/2041-8213/acade4},
archivePrefix = {arXiv},
       eprint = {2211.05792},
 primaryClass = {astro-ph.GA},
       adsurl = {https://ui.adsabs.harvard.edu/abs/2023ApJ...946L..13F}
}

@ARTICLE{javier23,
       author = {{{\'A}lvarez-M{\'a}rquez}, J. and {Crespo G{\'o}mez}, A. and {Colina}, L. and {Neeleman}, M. and {Walter}, F. and {Labiano}, A. and {P{\'e}rez-Gonz{\'a}lez}, P. and {Bik}, A. and {Noorgaard-Nielsen}, H.~U. and {Ostlin}, G. and {Wright}, G. and {Alonso-Herrero}, A. and {Azollini}, R. and {Caputi}, K.~I. and {Eckart}, A. and {Le F{\`e}vre}, O. and {Garc{\'\i}a-Mar{\'\i}n}, M. and {Greve}, T.~R. and {Hjorth}, J. and {Ilbert}, O. and {Kendrew}, S. and {Pye}, J.~P. and {Tikkanen}, T. and {Topinka}, M. and {van der Werf}, P. and {Ward}, M. and {van Dishoeck}, E.~F. and {G{\"u}del}, M. and {Henning}, Th. and {Lagage}, P.~O. and {Ray}, T. and {Waelkens}, C.},
        title = "{MIRI/JWST observations reveal an extremely obscured starburst in the z = 6.9 system SPT0311-58}",
      journal = {\aap},
         year = 2023,
        month = mar,
       volume = {671},
          eid = {A105},
        pages = {A105},
          doi = {10.1051/0004-6361/202245400},
archivePrefix = {arXiv},
       eprint = {2301.02313},
 primaryClass = {astro-ph.GA},
       adsurl = {https://ui.adsabs.harvard.edu/abs/2023A&A...671A.105A}
}

@ARTICLE{gomezguijarro2023,
       author = {{G{\'o}mez-Guijarro}, Carlos and {Magnelli}, Benjamin and {Elbaz}, David and {Wuyts}, Stijn and {Daddi}, Emanuele and {Le Bail}, Aur{\'e}lien and {Giavalisco}, Mauro and {Dickinson}, Mark and {P{\'e}rez-Gonz{\'a}lez}, Pablo G. and {Arrabal Haro}, Pablo and {Bagley}, Micaela B. and {Bisigello}, Laura and {Buat}, V{\'e}ronique and {Burgarella}, Denis and {Calabr{\`o}}, Antonello and {Casey}, Caitlin M. and {Cheng}, Yingjie and {Ciesla}, Laure and {Dekel}, Avishai and {Ferguson}, Henry C. and {Finkelstein}, Steven L. and {Franco}, Maximilien and {Grogin}, Norman A. and {Holwerda}, Benne W. and {Jin}, Shuowen and {Kartaltepe}, Jeyhan S. and {Koekemoer}, Anton M. and {Kokorev}, Vasily and {Long}, Arianna S. and {Lucas}, Ray A. and {Magdis}, Georgios E. and {Papovich}, Casey and {Pirzkal}, Nor and {Seill{\'e}}, Lise-Marie and {Tacchella}, Sandro and {Tarrasse}, Maxime and {Valentino}, Francesco and {de la Vega}, Alexander and {Wilkins}, Stephen M. and {Xiao}, Mengyuan and {Yung}, L.~Y. Aaron},
        title = "{JWST CEERS probes the role of stellar mass and morphology in obscuring galaxies}",
      journal = {\aap},
         year = 2023,
        month = sep,
       volume = {677},
          eid = {A34},
        pages = {A34},
          doi = {10.1051/0004-6361/202346673},
archivePrefix = {arXiv},
       eprint = {2304.08517},
 primaryClass = {astro-ph.GA},
       adsurl = {https://ui.adsabs.harvard.edu/abs/2023A&A...677A..34G}
}

@ARTICLE{Navarro-Carrera2024,
       author = {{Navarro-Carrera}, Rafael and {Rinaldi}, Pierluigi and {Caputi}, Karina I. and {Iani}, Edoardo and {Kokorev}, Vasily and {van Mierlo}, Sophie E.},
        title = "{Constraints on the Faint End of the Galaxy Stellar Mass Function at z ≃ 4{\textendash}8 from Deep JWST Data}",
      journal = {\apj},
         year = 2024,
        month = feb,
       volume = {961},
       number = {2},
          eid = {207},
        pages = {207},
          doi = {10.3847/1538-4357/ad0df6},
archivePrefix = {arXiv},
       eprint = {2305.16141},
 primaryClass = {astro-ph.GA},
       adsurl = {https://ui.adsabs.harvard.edu/abs/2024ApJ...961..207N}
}

@ARTICLE{Martorano2025,
       author = {{Martorano}, M. and {van der Wel}, A. and {Baes}, M. and {Bell}, E.~F. and {Brammer}, G. and {Franx}, M. and {Gebek}, A. and {Meidt}, S.~E. and {Miller}, T.~B. and {Nelson}, E. and {Nersesian}, A. and {Price}, S.~H. and {van Dokkum}, P. and {Whitaker}, K.~E. and {Wuyts}, S.},
        title = "{Evolution of the S{\'e}rsic index up to z = 2.5 from JWST and HST}",
      journal = {\aap},
         year = 2025,
        month = feb,
       volume = {694},
          eid = {A76},
        pages = {A76},
          doi = {10.1051/0004-6361/202452919},
archivePrefix = {arXiv},
       eprint = {2501.02956},
 primaryClass = {astro-ph.GA},
       adsurl = {https://ui.adsabs.harvard.edu/abs/2025A&A...694A..76M}
}

@ARTICLE{Weibel2024,
       author = {{Weibel}, Andrea and {Oesch}, Pascal A. and {Barrufet}, Laia and {Gottumukkala}, Rashmi and {Ellis}, Richard S. and {Santini}, Paola and {Weaver}, John R. and {Allen}, Natalie and {Bouwens}, Rychard and {Bowler}, Rebecca A.~A. and {Brammer}, Gabe and {Carnall}, Adam C. and {Cullen}, Fergus and {Dayal}, Pratika and {Dickinson}, Mark and {Donnan}, Callum T. and {Dunlop}, James S. and {Giavalisco}, Mauro and {Grogin}, Norman A. and {Illingworth}, Garth D. and {Koekemoer}, Anton M. and {Labbe}, Ivo and {Marchesini}, Danilo and {McLeod}, Derek J. and {McLure}, Ross J. and {Naidu}, Rohan P. and {P{\'e}rez-Gonz{\'a}lez}, Pablo G. and {Shuntov}, Marko and {Stefanon}, Mauro and {Toft}, Sune and {Xiao}, Mengyuan},
        title = "{Galaxy build-up in the first 1.5 Gyr of cosmic history: insights from the stellar mass function at z   4-9 from JWST NIRCam observations}",
      journal = {\mnras},
         year = 2024,
        month = sep,
       volume = {533},
       number = {2},
        pages = {1808-1838},
          doi = {10.1093/mnras/stae1891},
archivePrefix = {arXiv},
       eprint = {2403.08872},
 primaryClass = {astro-ph.GA},
       adsurl = {https://ui.adsabs.harvard.edu/abs/2024MNRAS.533.1808W}
}

@ARTICLE{Iani2024,
       author = {{Iani}, Edoardo and {Caputi}, Karina I. and {Rinaldi}, Pierluigi and {Annunziatella}, Marianna and {Boogaard}, Leindert A. and {{\"O}stlin}, G{\"o}ran and {Costantin}, Luca and {Gillman}, Steven and {P{\'e}rez-Gonz{\'a}lez}, Pablo G. and {Colina}, Luis and {Greve}, Thomas R. and {Wright}, Gillian and {Alonso-Herrero}, Almudena and {{\'A}lvarez-M{\'a}rquez}, Javier and {Bik}, Arjan and {Bosman}, Sarah E.~I. and {Crespo G{\'o}mez}, Alejandro and {Eckart}, Andreas and {Hjorth}, Jens and {Jermann}, Iris and {Labiano}, Alvaro and {Langeroodi}, Danial and {Melinder}, Jens and {Moutard}, Thibaud and {Pei{\ss}ker}, Florian and {Pye}, John P. and {Tikkanen}, Tuomo V. and {van der Werf}, Paul P. and {Walter}, Fabian and {Henning}, Thomas K. and {Lagage}, Pierre-Olivier and {van Dishoeck}, Ewine F.},
        title = "{MIDIS: JWST NIRCam and MIRI Unveil the Stellar Population Properties of Ly{\ensuremath{\alpha}} Emitters and Lyman-break Galaxies at z ≃ 3{\textendash}7}",
      journal = {\apj},
         year = 2024,
        month = mar,
       volume = {963},
       number = {2},
          eid = {97},
        pages = {97},
          doi = {10.3847/1538-4357/ad15f6},
archivePrefix = {arXiv},
       eprint = {2309.08515},
 primaryClass = {astro-ph.GA},
       adsurl = {https://ui.adsabs.harvard.edu/abs/2024ApJ...963...97I}
}

@ARTICLE{Li2024,
       author = {{Li}, Juno and {Da Cunha}, Elisabete and {Gonz{\'a}lez-L{\'o}pez}, Jorge and {Aravena}, Manuel and {De Looze}, Ilse and {F{\"o}rster Schreiber}, N.~M. and {Herrera-Camus}, Rodrigo and {Spilker}, Justin and {Tadaki}, Ken-ichi and {Barcos-Munoz}, Loreto and {Battisti}, Andrew J. and {Birkin}, Jack E. and {Bowler}, Rebecca A.~A. and {Davies}, Rebecca and {D{\'\i}az-Santos}, Tanio and {Ferrara}, Andrea and {Fisher}, Deanne B. and {Hodge}, Jacqueline and {Ikeda}, Ryota and {Killi}, Meghana and {Lee}, Lilian and {Liu}, Daizhong and {Lutz}, Dieter and {Mitsuhashi}, Ikki and {Naab}, Thorsten and {Posses}, Ana and {Rela{\~n}o}, Monica and {Solimano}, Manuel and {{\"U}bler}, Hannah and {van der Giessen}, Stefan Anthony and {Villanueva}, Vicente},
        title = "{The ALMA-CRISTAL Survey: Spatially Resolved Star Formation Activity and Dust Content in 4 < z < 6 Star-forming Galaxies}",
      journal = {\apj},
         year = 2024,
        month = nov,
       volume = {976},
       number = {1},
          eid = {70},
        pages = {70},
          doi = {10.3847/1538-4357/ad7fee},
archivePrefix = {arXiv},
       eprint = {2409.10961},
 primaryClass = {astro-ph.GA},
       adsurl = {https://ui.adsabs.harvard.edu/abs/2024ApJ...976...70L}
}

@ARTICLE{Wang2024,
       author = {{Wang}, Bingjie and {Leja}, Joel and {Labb{\'e}}, Ivo and {Bezanson}, Rachel and {Whitaker}, Katherine E. and {Brammer}, Gabriel and {Furtak}, Lukas J. and {Weaver}, John R. and {Price}, Sedona H. and {Zitrin}, Adi and {Atek}, Hakim and {Coe}, Dan and {Cutler}, Sam E. and {Dayal}, Pratika and {van Dokkum}, Pieter and {Feldmann}, Robert and {Marchesini}, Danilo and {Franx}, Marijn and {F{\"o}rster Schreiber}, Natascha and {Fujimoto}, Seiji and {Geha}, Marla and {Glazebrook}, Karl and {de Graaff}, Anna and {Greene}, Jenny E. and {Juneau}, St{\'e}phanie and {Kassin}, Susan and {Kriek}, Mariska and {Khullar}, Gourav and {Maseda}, Michael and {Mowla}, Lamiya A. and {Muzzin}, Adam and {Nanayakkara}, Themiya and {Nelson}, Erica J. and {Oesch}, Pascal A. and {Pacifici}, Camilla and {Pan}, Richard and {Papovich}, Casey and {Setton}, David J. and {Shapley}, Alice E. and {Smit}, Renske and {Stefanon}, Mauro and {Suess}, Katherine A. and {Taylor}, Edward N. and {Williams}, Christina C.},
        title = "{The UNCOVER Survey: A First-look HST+JWST Catalog of Galaxy Redshifts and Stellar Population Properties Spanning 0.2 {\ensuremath{\lesssim}} z {\ensuremath{\lesssim}} 15}",
      journal = {\apjs},
         year = 2024,
        month = jan,
       volume = {270},
       number = {1},
          eid = {12},
        pages = {12},
          doi = {10.3847/1538-4365/ad0846},
archivePrefix = {arXiv},
       eprint = {2310.01276},
 primaryClass = {astro-ph.GA},
       adsurl = {https://ui.adsabs.harvard.edu/abs/2024ApJS..270...12W}
}

@ARTICLE{Wuyts2013,
       author = {{Wuyts}, Stijn and {F{\"o}rster Schreiber}, Natascha M. and {Nelson}, Erica J. and {van Dokkum}, Pieter G. and {Brammer}, Gabe and {Chang}, Yu-Yen and {Faber}, Sandra M. and {Ferguson}, Henry C. and {Franx}, Marijn and {Fumagalli}, Mattia and {Genzel}, Reinhard and {Grogin}, Norman A. and {Kocevski}, Dale D. and {Koekemoer}, Anton M. and {Lundgren}, Britt and {Lutz}, Dieter and {McGrath}, Elizabeth J. and {Momcheva}, Ivelina and {Rosario}, David and {Skelton}, Rosalind E. and {Tacconi}, Linda J. and {van der Wel}, Arjen and {Whitaker}, Katherine E.},
        title = "{A CANDELS-3D-HST synergy: Resolved Star Formation Patterns at 0.7 < z < 1.5}",
      journal = {\apj},
         year = 2013,
        month = dec,
       volume = {779},
       number = {2},
          eid = {135},
        pages = {135},
          doi = {10.1088/0004-637X/779/2/135},
archivePrefix = {arXiv},
       eprint = {1310.5702},
 primaryClass = {astro-ph.CO},
       adsurl = {https://ui.adsabs.harvard.edu/abs/2013ApJ...779..135W}
}

@ARTICLE{Matharu2024,
       author = {{Matharu}, Jasleen and {Nelson}, Erica J. and {Brammer}, Gabriel and {Oesch}, Pascal A. and {Allen}, Natalie and {Shivaei}, Irene and {Naidu}, Rohan P. and {Chisholm}, John and {Covelo-Paz}, Alba and {Fudamoto}, Yoshinobu and {Giovinazzo}, Emma and {Herard-Demanche}, Thomas and {Kerutt}, Josephine and {Kramarenko}, Ivan and {Marchesini}, Danilo and {Meyer}, Romain A. and {Prieto-Lyon}, Gonzalo and {Reddy}, Naveen and {Shuntov}, Marko and {Weibel}, Andrea and {Wuyts}, Stijn and {Xiao}, Mengyuan},
        title = "{A first look at spatially resolved star formation at 4.8 < z < 6.5 with JWST FRESCO NIRCam slitless spectroscopy}",
      journal = {\aap},
         year = 2024,
        month = oct,
       volume = {690},
          eid = {A64},
        pages = {A64},
          doi = {10.1051/0004-6361/202450522},
archivePrefix = {arXiv},
       eprint = {2404.17629},
 primaryClass = {astro-ph.GA},
       adsurl = {https://ui.adsabs.harvard.edu/abs/2024A&A...690A..64M}
}

@ARTICLE{Lee2022,
       author = {{Lee}, Janice C. and {Whitmore}, Bradley C. and {Thilker}, David A. and {Deger}, Sinan and {Larson}, Kirsten L. and {Ubeda}, Leonardo and {Anand}, Gagandeep S. and {Boquien}, M{\'e}d{\'e}ric and {Chandar}, Rupali and {Dale}, Daniel A. and {Emsellem}, Eric and {Leroy}, Adam K. and {Rosolowsky}, Erik and {Schinnerer}, Eva and {Schmidt}, Judy and {Lilly}, James and {Turner}, Jordan and {Van Dyk}, Schuyler and {White}, Richard L. and {Barnes}, Ashley T. and {Belfiore}, Francesco and {Bigiel}, Frank and {Blanc}, Guillermo A. and {Cao}, Yixian and {Chevance}, Melanie and {Congiu}, Enrico and {Egorov}, Oleg V. and {Glover}, Simon C.~O. and {Grasha}, Kathryn and {Groves}, Brent and {Henshaw}, Jonathan D. and {Hughes}, Annie and {Klessen}, Ralf S. and {Koch}, Eric and {Kreckel}, Kathryn and {Kruijssen}, J.~M. Diederik and {Liu}, Daizhong and {Lopez}, Laura A. and {Mayker}, Ness and {Meidt}, Sharon E. and {Murphy}, Eric J. and {Pan}, Hsi-An and {Pety}, J{\'e}r{\^o}me and {Querejeta}, Miguel and {Razza}, Alessandro and {Saito}, Toshiki and {S{\'a}nchez-Bl{\'a}zquez}, Patricia and {Santoro}, Francesco and {Sardone}, Amy and {Scheuermann}, Fabian and {Schruba}, Andreas and {Sun}, Jiayi and {Usero}, Antonio and {Watkins}, E. and {Williams}, Thomas G.},
        title = "{The PHANGS-HST Survey: Physics at High Angular Resolution in Nearby Galaxies with the Hubble Space Telescope}",
      journal = {\apjs},
         year = 2022,
        month = jan,
       volume = {258},
       number = {1},
          eid = {10},
        pages = {10},
          doi = {10.3847/1538-4365/ac1fe5},
archivePrefix = {arXiv},
       eprint = {2101.02855},
 primaryClass = {astro-ph.GA},
       adsurl = {https://ui.adsabs.harvard.edu/abs/2022ApJS..258...10L}
}

@ARTICLE{Song2023,
       author = {{Song}, Jie and {Fang}, GuanWen and {Lin}, Zesen and {Gu}, Yizhou and {Kong}, Xu},
        title = "{Solution to the Conflict between the Estimations of Resolved and Unresolved Galaxy Stellar Mass from the Perspective of JWST}",
      journal = {\apj},
         year = 2023,
        month = nov,
       volume = {958},
       number = {1},
          eid = {82},
        pages = {82},
          doi = {10.3847/1538-4357/ad0365},
archivePrefix = {arXiv},
       eprint = {2310.12228},
 primaryClass = {astro-ph.GA},
       adsurl = {https://ui.adsabs.harvard.edu/abs/2023ApJ...958...82S}
}

@ARTICLE{Arteaga2023,
       author = {{Gim{\'e}nez-Arteaga}, Clara and {Oesch}, Pascal A. and {Brammer}, Gabriel B. and {Valentino}, Francesco and {Mason}, Charlotte A. and {Weibel}, Andrea and {Barrufet}, Laia and {Fujimoto}, Seiji and {Heintz}, Kasper E. and {Nelson}, Erica J. and {Strait}, Victoria B. and {Suess}, Katherine A. and {Gibson}, Justus},
        title = "{Spatially Resolved Properties of Galaxies at 5 < z < 9 in the SMACS 0723 JWST ERO Field}",
      journal = {\apj},
         year = 2023,
        month = may,
       volume = {948},
       number = {2},
          eid = {126},
        pages = {126},
          doi = {10.3847/1538-4357/acc5ea},
archivePrefix = {arXiv},
       eprint = {2212.08670},
 primaryClass = {astro-ph.GA},
       adsurl = {https://ui.adsabs.harvard.edu/abs/2023ApJ...948..126G}
}

@ARTICLE{Wang2022,
       author = {{Wang}, Xin and {Jones}, Tucker and {Vulcani}, Benedetta and {Treu}, Tommaso and {Morishita}, Takahiro and {Roberts-Borsani}, Guido and {Malkan}, Matthew A. and {Henry}, Alaina and {Brammer}, Gabriel and {Strait}, Victoria and {Brada{\v{c}}}, Maru{\v{s}}a and {Boyett}, Kristan and {Calabr{\`o}}, Antonello and {Castellano}, Marco and {Fontana}, Adriano and {Glazebrook}, Karl and {Kelly}, Patrick L. and {Leethochawalit}, Nicha and {Marchesini}, Danilo and {Santini}, P. and {Trenti}, M. and {Yang}, Lilan},
        title = "{Early Results from GLASS-JWST. IV. Spatially Resolved Metallicity in a Low-mass z   3 Galaxy with NIRISS}",
      journal = {\apjl},
         year = 2022,
        month = oct,
       volume = {938},
       number = {2},
          eid = {L16},
        pages = {L16},
          doi = {10.3847/2041-8213/ac959e},
archivePrefix = {arXiv},
       eprint = {2207.13113},
 primaryClass = {astro-ph.GA},
       adsurl = {https://ui.adsabs.harvard.edu/abs/2022ApJ...938L..16W}
}

@ARTICLE{pg2023,
       author = {{P{\'e}rez-Gonz{\'a}lez}, Pablo G. and {Barro}, Guillermo and {Annunziatella}, Marianna and {Costantin}, Luca and {Garc{\'\i}a-Argum{\'a}nez}, {\'A}ngela and {McGrath}, Elizabeth J. and {M{\'e}rida}, Rosa M. and {Zavala}, Jorge A. and {Arrabal Haro}, Pablo and {Bagley}, Micaela B. and {Backhaus}, Bren E. and {Behroozi}, Peter and {Bell}, Eric F. and {Bisigello}, Laura and {Buat}, V{\'e}ronique and {Calabr{\`o}}, Antonello and {Casey}, Caitlin M. and {Cleri}, Nikko J. and {Coogan}, Rosemary T. and {Cooper}, M.~C. and {Cooray}, Asantha R. and {Dekel}, Avishai and {Dickinson}, Mark and {Elbaz}, David and {Ferguson}, Henry C. and {Finkelstein}, Steven L. and {Fontana}, Adriano and {Franco}, Maximilien and {Gardner}, Jonathan P. and {Giavalisco}, Mauro and {G{\'o}mez-Guijarro}, Carlos and {Grazian}, Andrea and {Grogin}, Norman A. and {Guo}, Yuchen and {Huertas-Company}, Marc and {Jogee}, Shardha and {Kartaltepe}, Jeyhan S. and {Kewley}, Lisa J. and {Kirkpatrick}, Allison and {Kocevski}, Dale D. and {Koekemoer}, Anton M. and {Long}, Arianna S. and {Lotz}, Jennifer M. and {Lucas}, Ray A. and {Papovich}, Casey and {Pirzkal}, Nor and {Ravindranath}, Swara and {Somerville}, Rachel S. and {Tacchella}, Sandro and {Trump}, Jonathan R. and {Wang}, Weichen and {Wilkins}, Stephen M. and {Wuyts}, Stijn and {Yang}, Guang and {Yung}, L.~Y. Aaron},
        title = "{CEERS Key Paper. IV. A Triality in the Nature of HST-dark Galaxies}",
      journal = {\apjl},
         year = 2023,
        month = mar,
       volume = {946},
       number = {1},
          eid = {L16},
        pages = {L16},
          doi = {10.3847/2041-8213/acb3a5},
archivePrefix = {arXiv},
       eprint = {2211.00045},
 primaryClass = {astro-ph.GA},
       adsurl = {https://ui.adsabs.harvard.edu/abs/2023ApJ...946L..16P}
}

@ARTICLE{Sorba2018,
       author = {{Sorba}, Robert and {Sawicki}, Marcin},
        title = "{Spatially unresolved SED fitting can underestimate galaxy masses: a solution to the missing mass problem}",
      journal = {\mnras},
         year = 2018,
        month = may,
       volume = {476},
       number = {2},
        pages = {1532-1547},
          doi = {10.1093/mnras/sty186},
archivePrefix = {arXiv},
       eprint = {1801.07368},
 primaryClass = {astro-ph.GA},
       adsurl = {https://ui.adsabs.harvard.edu/abs/2018MNRAS.476.1532S}
}

@ARTICLE{Zibetti2009,
       author = {{Zibetti}, Stefano and {Charlot}, St{\'e}phane and {Rix}, Hans-Walter},
        title = "{Resolved stellar mass maps of galaxies - I. Method and implications for global mass estimates}",
      journal = {\mnras},
         year = 2009,
        month = dec,
       volume = {400},
       number = {3},
        pages = {1181-1198},
          doi = {10.1111/j.1365-2966.2009.15528.x},
archivePrefix = {arXiv},
       eprint = {0904.4252},
 primaryClass = {astro-ph.CO},
       adsurl = {https://ui.adsabs.harvard.edu/abs/2009MNRAS.400.1181Z}
}

@ARTICLE{Sorba2015,
       author = {{Sorba}, R. and {Sawicki}, M.},
        title = "{Missing stellar mass in SED fitting: spatially unresolved photometry can underestimate galaxy masses}",
      journal = {\mnras},
         year = 2015,
        month = sep,
       volume = {452},
       number = {1},
        pages = {235-245},
          doi = {10.1093/mnras/stv1235},
archivePrefix = {arXiv},
       eprint = {1506.01653},
 primaryClass = {astro-ph.GA},
       adsurl = {https://ui.adsabs.harvard.edu/abs/2015MNRAS.452..235S}
}

@ARTICLE{arteaga2024,
       author = {{Gim{\'e}nez-Arteaga}, C. and {Fujimoto}, S. and {Valentino}, F. and {Brammer}, G.~B. and {Mason}, C.~A. and {Rizzo}, F. and {Rusakov}, V. and {Colina}, L. and {Prieto-Lyon}, G. and {Oesch}, P.~A. and {Espada}, D. and {Heintz}, K.~E. and {Knudsen}, K.~K. and {Dessauges-Zavadsky}, M. and {Laporte}, N. and {Lee}, M. and {Magdis}, G.~E. and {Ono}, Y. and {Ao}, Y. and {Ouchi}, M. and {Kohno}, K. and {Koekemoer}, A.~M.},
        title = "{Outshining in the spatially resolved analysis of a strongly lensed galaxy at z = 6.072 with JWST NIRCam}",
      journal = {\aap},
         year = 2024,
        month = jun,
       volume = {686},
          eid = {A63},
        pages = {A63},
          doi = {10.1051/0004-6361/202349135},
archivePrefix = {arXiv},
       eprint = {2402.17875},
 primaryClass = {astro-ph.GA},
       adsurl = {https://ui.adsabs.harvard.edu/abs/2024A&A...686A..63G}
}

@ARTICLE{Narayanan2024,
       author = {{Narayanan}, Desika and {Lower}, Sidney and {Torrey}, Paul and {Brammer}, Gabriel and {Cui}, Weiguang and {Dav{\'e}}, Romeel and {Iyer}, Kartheik G. and {Li}, Qi and {Lovell}, Christopher C. and {Sales}, Laura V. and {Stark}, Daniel P. and {Marinacci}, Federico and {Vogelsberger}, Mark},
        title = "{Outshining by Recent Star Formation Prevents the Accurate Measurement of High-z Galaxy Stellar Masses}",
      journal = {\apj},
         year = 2024,
        month = jan,
       volume = {961},
       number = {1},
          eid = {73},
        pages = {73},
          doi = {10.3847/1538-4357/ad0966},
archivePrefix = {arXiv},
       eprint = {2306.10118},
 primaryClass = {astro-ph.GA},
       adsurl = {https://ui.adsabs.harvard.edu/abs/2024ApJ...961...73N}
}

@ARTICLE{Papovich2001,
       author = {{Papovich}, Casey and {Dickinson}, Mark and {Ferguson}, Henry C.},
        title = "{The Stellar Populations and Evolution of Lyman Break Galaxies}",
      journal = {\apj},
         year = 2001,
        month = oct,
       volume = {559},
       number = {2},
        pages = {620-653},
          doi = {10.1086/322412},
archivePrefix = {arXiv},
       eprint = {astro-ph/0105087},
 primaryClass = {astro-ph},
       adsurl = {https://ui.adsabs.harvard.edu/abs/2001ApJ...559..620P}
}

@ARTICLE{Tacchella2022,
       author = {{Tacchella}, Sandro and {Finkelstein}, Steven L. and {Bagley}, Micaela and {Dickinson}, Mark and {Ferguson}, Henry C. and {Giavalisco}, Mauro and {Graziani}, Luca and {Grogin}, Norman A. and {Hathi}, Nimish and {Hutchison}, Taylor A. and {Jung}, Intae and {Koekemoer}, Anton M. and {Larson}, Rebecca L. and {Papovich}, Casey and {Pirzkal}, Norbert and {Rojas-Ruiz}, Sof{\'\i}a and {Song}, Mimi and {Schneider}, Raffaella and {Somerville}, Rachel S. and {Wilkins}, Stephen M. and {Yung}, L.~Y. Aaron},
        title = "{On the Stellar Populations of Galaxies at z = 9-11: The Growth of Metals and Stellar Mass at Early Times}",
      journal = {\apj},
         year = 2022,
        month = mar,
       volume = {927},
       number = {2},
          eid = {170},
        pages = {170},
          doi = {10.3847/1538-4357/ac4cad},
archivePrefix = {arXiv},
       eprint = {2111.05351},
 primaryClass = {astro-ph.GA},
       adsurl = {https://ui.adsabs.harvard.edu/abs/2022ApJ...927..170T}
}

@ARTICLE{Topping2022,
       author = {{Topping}, Michael W. and {Stark}, Daniel P. and {Endsley}, Ryan and {Bouwens}, Rychard J. and {Schouws}, Sander and {Smit}, Renske and {Stefanon}, Mauro and {Inami}, Hanae and {Bowler}, Rebecca A.~A. and {Oesch}, Pascal and {Gonzalez}, Valentino and {Dayal}, Pratika and {da Cunha}, Elisabete and {Algera}, Hiddo and {van der Werf}, Paul and {Pallottini}, Andrea and {Barrufet}, Laia and {Schneider}, Raffaella and {De Looze}, Ilse and {Sommovigo}, Laura and {Whitler}, Lily and {Graziani}, Luca and {Fudamoto}, Yoshinobu and {Ferrara}, Andrea},
        title = "{The ALMA REBELS Survey: specific star formation rates in the reionization era}",
      journal = {\mnras},
         year = 2022,
        month = oct,
       volume = {516},
       number = {1},
        pages = {975-991},
          doi = {10.1093/mnras/stac2291},
archivePrefix = {arXiv},
       eprint = {2203.07392},
 primaryClass = {astro-ph.GA},
       adsurl = {https://ui.adsabs.harvard.edu/abs/2022MNRAS.516..975T}
}

@ARTICLE{Whitler2023,
       author = {{Whitler}, Lily and {Stark}, Daniel P. and {Endsley}, Ryan and {Leja}, Joel and {Charlot}, St{\'e}phane and {Chevallard}, Jacopo},
        title = "{Star formation histories of UV-luminous galaxies at z ≃ 6.8: implications for stellar mass assembly at early cosmic times}",
      journal = {\mnras},
         year = 2023,
        month = mar,
       volume = {519},
       number = {4},
        pages = {5859-5881},
          doi = {10.1093/mnras/stad004},
archivePrefix = {arXiv},
       eprint = {2206.05315},
 primaryClass = {astro-ph.GA},
       adsurl = {https://ui.adsabs.harvard.edu/abs/2023MNRAS.519.5859W}
}

@ARTICLE{Shen2024,
       author = {{Shen}, Lu and {Papovich}, Casey and {Matharu}, Jasleen and {Pirzkal}, Nor and {Hu}, Weida and {Backhaus}, Bren E. and {Bagley}, Micaela B. and {Cheng}, Yingjie and {Cleri}, Nikko J. and {Finkelstein}, Steven L. and {Huertas-Company}, Marc and {Giavalisco}, Mauro and {Grogin}, Norman A. and {Jung}, Intae and {Kartaltepe}, Jeyhan S. and {Koekemoer}, Anton M. and {Lotz}, Jennifer M. and {Maseda}, Michael V. and {P{\'e}rez-Gonz{\'a}lez}, Pablo G. and {Rothberg}, Barry and {Simons}, Raymond C. and {Tacchella}, Sandro and {Williams}, Christina C. and {Yung}, L.~Y. Aaron},
        title = "{NGDEEP Epoch 1: Spatially Resolved H{\ensuremath{\alpha}} Observations of Disk and Bulge Growth in Star-forming Galaxies at z {\ensuremath{\sim}} 0.6{\textendash}2.2 from JWST NIRISS Slitless Spectroscopy}",
      journal = {\apjl},
         year = 2024,
        month = mar,
       volume = {963},
       number = {2},
          eid = {L49},
        pages = {L49},
          doi = {10.3847/2041-8213/ad28bd},
archivePrefix = {arXiv},
       eprint = {2310.13745},
 primaryClass = {astro-ph.GA},
       adsurl = {https://ui.adsabs.harvard.edu/abs/2024ApJ...963L..49S}
}

@ARTICLE{Witten2025,
       author = {{Witten}, Callum and {McClymont}, William and {Laporte}, Nicolas and {Roberts-Borsani}, Guido and {Sijacki}, Debora and {Tacchella}, Sandro and {Simmonds}, Charlotte and {Katz}, Harley and {Ellis}, Richard S. and {Witstok}, Joris and {Maiolino}, Roberto and {Ji}, Xihan and {Hayes}, Billy R. and {Looser}, Tobias J. and {D'Eugenio}, Francesco},
        title = "{Rising from the ashes: evidence of old stellar populations and rejuvenation events in the very early Universe}",
      journal = {\mnras},
         year = 2025,
        month = feb,
       volume = {537},
       number = {1},
        pages = {112-126},
          doi = {10.1093/mnras/staf001},
archivePrefix = {arXiv},
       eprint = {2407.07937},
 primaryClass = {astro-ph.GA},
       adsurl = {https://ui.adsabs.harvard.edu/abs/2025MNRAS.537..112W}
}

@ARTICLE{Cibinel2015,
       author = {{Cibinel}, A. and {Le Floc'h}, E. and {Perret}, V. and {Bournaud}, F. and {Daddi}, E. and {Pannella}, M. and {Elbaz}, D. and {Amram}, P. and {Duc}, P. -A.},
        title = "{A Physical Approach to the Identification of High-z Mergers: Morphological Classification in the Stellar Mass Domain.}",
      journal = {\apj},
         year = 2015,
        month = jun,
       volume = {805},
       number = {2},
          eid = {181},
        pages = {181},
          doi = {10.1088/0004-637X/805/2/181},
archivePrefix = {arXiv},
       eprint = {1503.06220},
 primaryClass = {astro-ph.GA},
       adsurl = {https://ui.adsabs.harvard.edu/abs/2015ApJ...805..181C}
}

@ARTICLE{Lines2025,
       author = {{Lines}, N.~E.~P. and {Bowler}, R.~A.~A. and {Adams}, N.~J. and {Fisher}, R. and {Varadaraj}, R.~G. and {Nakazato}, Y. and {Aravena}, M. and {Assef}, R.~J. and {Birkin}, J.~E. and {Ceverino}, D. and {da Cunha}, E. and {Cullen}, F. and {De Looze}, I. and {Donnan}, C.~T. and {Dunlop}, J.~S. and {Ferrara}, A. and {Grogin}, N.~A. and {Herrera-Camus}, R. and {Ikeda}, R. and {Koekemoer}, A.~M. and {Killi}, M. and {Li}, J. and {McLeod}, D.~J. and {McLure}, R.~J. and {Mitsuhashi}, I. and {P{\'e}rez-Gonz{\'a}lez}, P.~G. and {Relano}, M. and {Solimano}, M. and {Spilker}, J.~S. and {Villanueva}, V. and {Yoshida}, N.},
        title = "{JWST PRIMER: a lack of outshining in four normal z = 4 ‑ 6 galaxies from the ALMA-CRISTAL Survey}",
      journal = {\mnras},
         year = 2025,
        month = may,
       volume = {539},
       number = {3},
        pages = {2685-2706},
          doi = {10.1093/mnras/staf627},
archivePrefix = {arXiv},
       eprint = {2409.10963},
 primaryClass = {astro-ph.GA},
       adsurl = {https://ui.adsabs.harvard.edu/abs/2025MNRAS.539.2685L}
}

@ARTICLE{Hemmati2014,
       author = {{Hemmati}, Shoubaneh and {Miller}, Sarah H. and {Mobasher}, Bahram and {Nayyeri}, Hooshang and {Ferguson}, Henry C. and {Guo}, Yicheng and {Koekemoer}, Anton M. and {Koo}, David C. and {Papovich}, Casey},
        title = "{Kiloparsec-scale Properties of Emission-line Galaxies}",
      journal = {\apj},
         year = 2014,
        month = dec,
       volume = {797},
       number = {2},
          eid = {108},
        pages = {108},
          doi = {10.1088/0004-637X/797/2/108},
archivePrefix = {arXiv},
       eprint = {1409.4791},
 primaryClass = {astro-ph.GA},
       adsurl = {https://ui.adsabs.harvard.edu/abs/2014ApJ...797..108H}
}

@ARTICLE{Pforr2012,
       author = {{Pforr}, Janine and {Maraston}, Claudia and {Tonini}, Chiara},
        title = "{Recovering galaxy stellar population properties from broad-band spectral energy distribution fitting}",
      journal = {\mnras},
         year = 2012,
        month = jun,
       volume = {422},
       number = {4},
        pages = {3285-3326},
          doi = {10.1111/j.1365-2966.2012.20848.x},
archivePrefix = {arXiv},
       eprint = {1203.3548},
 primaryClass = {astro-ph.CO},
       adsurl = {https://ui.adsabs.harvard.edu/abs/2012MNRAS.422.3285P}
}

@ARTICLE{Gallazzi2009,
       author = {{Gallazzi}, Anna and {Bell}, Eric F.},
        title = "{Stellar Mass-to-Light Ratios from Galaxy Spectra: How Accurate Can They Be?}",
      journal = {\apjs},
         year = 2009,
        month = dec,
       volume = {185},
       number = {2},
        pages = {253-272},
          doi = {10.1088/0067-0049/185/2/253},
archivePrefix = {arXiv},
       eprint = {0910.1591},
 primaryClass = {astro-ph.CO},
       adsurl = {https://ui.adsabs.harvard.edu/abs/2009ApJS..185..253G}
}

@ARTICLE{Bolzonella2010,
       author = {{Bolzonella}, M. and {Kova{\v{c}}}, K. and {Pozzetti}, L. and {Zucca}, E. and {Cucciati}, O. and {Lilly}, S.~J. and {Peng}, Y. and {Iovino}, A. and {Zamorani}, G. and {Vergani}, D. and {Tasca}, L.~A.~M. and {Lamareille}, F. and {Oesch}, P. and {Caputi}, K. and {Kampczyk}, P. and {Bardelli}, S. and {Maier}, C. and {Abbas}, U. and {Knobel}, C. and {Scodeggio}, M. and {Carollo}, C.~M. and {Contini}, T. and {Kneib}, J. -P. and {Le F{\`e}vre}, O. and {Mainieri}, V. and {Renzini}, A. and {Bongiorno}, A. and {Coppa}, G. and {de la Torre}, S. and {de Ravel}, L. and {Franzetti}, P. and {Garilli}, B. and {Le Borgne}, J. -F. and {Le Brun}, V. and {Mignoli}, M. and {Pell{\'o}}, R. and {Perez-Montero}, E. and {Ricciardelli}, E. and {Silverman}, J.~D. and {Tanaka}, M. and {Tresse}, L. and {Bottini}, D. and {Cappi}, A. and {Cassata}, P. and {Cimatti}, A. and {Guzzo}, L. and {Koekemoer}, A.~M. and {Leauthaud}, A. and {Maccagni}, D. and {Marinoni}, C. and {McCracken}, H.~J. and {Memeo}, P. and {Meneux}, B. and {Porciani}, C. and {Scaramella}, R. and {Aussel}, H. and {Capak}, P. and {Halliday}, C. and {Ilbert}, O. and {Kartaltepe}, J. and {Salvato}, M. and {Sanders}, D. and {Scarlata}, C. and {Scoville}, N. and {Taniguchi}, Y. and {Thompson}, D.},
        title = "{Tracking the impact of environment on the galaxy stellar mass function up to z \raisebox{-0.5ex}\textasciitilde 1 in the 10 k zCOSMOS sample}",
      journal = {\aap},
         year = 2010,
        month = dec,
       volume = {524},
          eid = {A76},
        pages = {A76},
          doi = {10.1051/0004-6361/200912801},
archivePrefix = {arXiv},
       eprint = {0907.0013},
 primaryClass = {astro-ph.CO},
       adsurl = {https://ui.adsabs.harvard.edu/abs/2010A&A...524A..76B}
}

@ARTICLE{Looser2025,
       author = {{Looser}, Tobias J. and {D'Eugenio}, Francesco and {Maiolino}, Roberto and {Tacchella}, Sandro and {Curti}, Mirko and {Arribas}, Santiago and {Baker}, William M. and {Baum}, Stefi and {Bonaventura}, Nina and {Boyett}, Kristan and {Bunker}, Andrew J. and {Carniani}, Stefano and {Charlot}, Stephane and {Chevallard}, Jacopo and {Curtis-Lake}, Emma and {Lola Danhaive}, A. and {Eisenstein}, Daniel J. and {de Graaff}, Anna and {Hainline}, Kevin and {Ji}, Zhiyuan and {Johnson}, Benjamin D. and {Kumari}, Nimisha and {Nelson}, Erica and {Parlanti}, Eleonora and {Rix}, Hans-Walter and {Robertson}, Brant and {Del Pino}, Bruno Rodr{\'\i}guez and {Sandles}, Lester and {Scholtz}, Jan and {Smit}, Renske and {Stark}, Daniel P. and {{\"U}bler}, Hannah and {Williams}, Christina C. and {Willott}, Chris and {Witstok}, Joris},
        title = "{JADES: Differing assembly histories of galaxies: Observational evidence for bursty star formation histories and (mini-)quenching in the first billion years of the Universe}",
      journal = {\aap},
         year = 2025,
        month = may,
       volume = {697},
          eid = {A88},
        pages = {A88},
          doi = {10.1051/0004-6361/202347102},
archivePrefix = {arXiv},
       eprint = {2306.02470},
 primaryClass = {astro-ph.GA},
       adsurl = {https://ui.adsabs.harvard.edu/abs/2025A&A...697A..88L}
}

@ARTICLE{Battisti2016,
       author = {{Battisti}, A.~J. and {Calzetti}, D. and {Chary}, R. -R.},
        title = "{Characterizing Dust Attenuation in Local Star-forming Galaxies: UV and Optical Reddening}",
      journal = {\apj},
         year = 2016,
        month = feb,
       volume = {818},
       number = {1},
          eid = {13},
        pages = {13},
          doi = {10.3847/0004-637X/818/1/13},
archivePrefix = {arXiv},
       eprint = {1601.00208},
 primaryClass = {astro-ph.GA},
       adsurl = {https://ui.adsabs.harvard.edu/abs/2016ApJ...818...13B}
}

@ARTICLE{Mascia2021,
       author = {{Di Mascia}, F. and {Gallerani}, S. and {Ferrara}, A. and {Pallottini}, A. and {Maiolino}, R. and {Carniani}, S. and {D'Odorico}, V.},
        title = "{The dust attenuation law in z   6 quasars}",
      journal = {\mnras},
         year = 2021,
        month = sep,
       volume = {506},
       number = {3},
        pages = {3946-3961},
          doi = {10.1093/mnras/stab1876},
archivePrefix = {arXiv},
       eprint = {2106.15625},
 primaryClass = {astro-ph.GA},
       adsurl = {https://ui.adsabs.harvard.edu/abs/2021MNRAS.506.3946D}
}

@ARTICLE{Merlin2024,
       author = {{Merlin}, E. and {Santini}, P. and {Paris}, D. and {Castellano}, M. and {Fontana}, A. and {Treu}, T. and {Finkelstein}, S.~L. and {Dunlop}, J.~S. and {Arrabal Haro}, P. and {Bagley}, M. and {Boyett}, K. and {Calabr{\`o}}, A. and {Correnti}, M. and {Davis}, K. and {Dickinson}, M. and {Donnan}, C.~T. and {Ferguson}, H.~C. and {Fortuni}, F. and {Giavalisco}, M. and {Glazebrook}, K. and {Grazian}, A. and {Grogin}, N.~A. and {Hathi}, N. and {Hirschmann}, M. and {Kartaltepe}, J.~S. and {Kewley}, L.~J. and {Kirkpatrick}, A. and {Kocevski}, D.~D. and {Koekemoer}, A.~M. and {Leung}, G. and {Lotz}, J.~M. and {Lucas}, R.~A. and {Magee}, D.~K. and {Marchesini}, D. and {Mascia}, S. and {McLeod}, D.~J. and {McLure}, R.~J. and {Nanayakkara}, T. and {Napolitano}, L. and {Nonino}, M. and {Papovich}, C. and {Pentericci}, L. and {P{\'e}rez-Gonz{\'a}lez}, P.~G. and {Pirzkal}, N. and {Ravindranath}, S. and {Roberts-Borsani}, G. and {Somerville}, R.~S. and {Trenti}, M. and {Trump}, J.~R. and {Vulcani}, B. and {Wang}, X. and {Watson}, P.~J. and {Wilkins}, S.~M. and {Yang}, G. and {Yung}, L.~Y.~A.},
        title = "{ASTRODEEP-JWST: NIRCam-HST multi-band photometry and redshifts for half a million sources in six extragalactic deep fields}",
      journal = {\aap},
         year = 2024,
        month = nov,
       volume = {691},
          eid = {A240},
        pages = {A240},
          doi = {10.1051/0004-6361/202451409},
archivePrefix = {arXiv},
       eprint = {2409.00169},
 primaryClass = {astro-ph.GA},
       adsurl = {https://ui.adsabs.harvard.edu/abs/2024A&A...691A.240M}
}

@ARTICLE{Giavalisco2004,
       author = {{Giavalisco}, M. and {Ferguson}, H.~C. and {Koekemoer}, A.~M. and {Dickinson}, M. and {Alexander}, D.~M. and {Bauer}, F.~E. and {Bergeron}, J. and {Biagetti}, C. and {Brandt}, W.~N. and {Casertano}, S. and {Cesarsky}, C. and {Chatzichristou}, E. and {Conselice}, C. and {Cristiani}, S. and {Da Costa}, L. and {Dahlen}, T. and {de Mello}, D. and {Eisenhardt}, P. and {Erben}, T. and {Fall}, S.~M. and {Fassnacht}, C. and {Fosbury}, R. and {Fruchter}, A. and {Gardner}, J.~P. and {Grogin}, N. and {Hook}, R.~N. and {Hornschemeier}, A.~E. and {Idzi}, R. and {Jogee}, S. and {Kretchmer}, C. and {Laidler}, V. and {Lee}, K.~S. and {Livio}, M. and {Lucas}, R. and {Madau}, P. and {Mobasher}, B. and {Moustakas}, L.~A. and {Nonino}, M. and {Padovani}, P. and {Papovich}, C. and {Park}, Y. and {Ravindranath}, S. and {Renzini}, A. and {Richardson}, M. and {Riess}, A. and {Rosati}, P. and {Schirmer}, M. and {Schreier}, E. and {Somerville}, R.~S. and {Spinrad}, H. and {Stern}, D. and {Stiavelli}, M. and {Strolger}, L. and {Urry}, C.~M. and {Vandame}, B. and {Williams}, R. and {Wolf}, C.},
        title = "{The Great Observatories Origins Deep Survey: Initial Results from Optical and Near-Infrared Imaging}",
      journal = {\apjl},
         year = 2004,
        month = jan,
       volume = {600},
       number = {2},
        pages = {L93-L98},
          doi = {10.1086/379232},
archivePrefix = {arXiv},
       eprint = {astro-ph/0309105},
 primaryClass = {astro-ph},
       adsurl = {https://ui.adsabs.harvard.edu/abs/2004ApJ...600L..93G}
}

@ARTICLE{Beckwith2006,
       author = {{Beckwith}, Steven V.~W. and {Stiavelli}, Massimo and {Koekemoer}, Anton M. and {Caldwell}, John A.~R. and {Ferguson}, Henry C. and {Hook}, Richard and {Lucas}, Ray A. and {Bergeron}, Louis E. and {Corbin}, Michael and {Jogee}, Shardha and {Panagia}, Nino and {Robberto}, Massimo and {Royle}, Patricia and {Somerville}, Rachel S. and {Sosey}, Megan},
        title = "{The Hubble Ultra Deep Field}",
      journal = {\aj},
         year = 2006,
        month = nov,
       volume = {132},
       number = {5},
        pages = {1729-1755},
          doi = {10.1086/507302},
archivePrefix = {arXiv},
       eprint = {astro-ph/0607632},
 primaryClass = {astro-ph},
       adsurl = {https://ui.adsabs.harvard.edu/abs/2006AJ....132.1729B}
}

@ARTICLE{Eisenstein2025,
       author = {{Eisenstein}, Daniel J. and {Johnson}, Benjamin D. and {Robertson}, Brant and {Tacchella}, Sandro and {Hainline}, Kevin and {Jakobsen}, Peter and {Maiolino}, Roberto and {Bonaventura}, Nina and {Bunker}, Andrew J. and {Cameron}, Alex J. and {Cargile}, Phillip A. and {Curtis-Lake}, Emma and {Hausen}, Ryan and {Pusk{\'a}s}, D{\'a}vid and {Rieke}, Marcia and {Sun}, Fengwu and {Willmer}, Christopher N.~A. and {Willott}, Chris and {Alberts}, Stacey and {Arribas}, Santiago and {Baker}, William M. and {Baum}, Stefi and {Bhatawdekar}, Rachana and {Carniani}, Stefano and {Charlot}, Stephane and {Chen}, Zuyi and {Chevallard}, Jacopo and {Curti}, Mirko and {DeCoursey}, Christa and {D'Eugenio}, Francesco and {de Graaff}, Anna and {Egami}, Eiichi and {Helton}, Jakob M. and {Ji}, Zhiyuan and {Jones}, Gareth C. and {Kumari}, Nimisha and {L{\"u}tzgendorf}, Nora and {Laseter}, Isaac and {Looser}, Tobias J. and {Lyu}, Jianwei and {Maseda}, Michael V. and {Nelson}, Erica and {Parlanti}, Eleonora and {Rauscher}, Bernard J. and {Rawle}, Tim and {Rieke}, George and {Rix}, Hans-Walter and {Rujopakarn}, Wiphu and {Sandles}, Lester and {Saxena}, Aayush and {Scholtz}, Jan and {Sharpe}, Katherine and {Shivaei}, Irene and {Simmonds}, Charlotte and {Smit}, Renske and {Topping}, Michael W. and {{\"U}bler}, Hannah and {Venturi}, Giacomo and {Williams}, Christina C. and {Witstok}, Joris and {Woodrum}, Charity},
        title = "{The JADES Origins Field: A New JWST Deep Field in the JADES Second NIRCam Data Release}",
      journal = {\apjs},
         year = 2025,
        month = dec,
       volume = {281},
       number = {2},
          eid = {50},
        pages = {50},
          doi = {10.3847/1538-4365/ae1137},
archivePrefix = {arXiv},
       eprint = {2310.12340},
 primaryClass = {astro-ph.GA},
       adsurl = {https://ui.adsabs.harvard.edu/abs/2025ApJS..281...50E}
}

@ARTICLE{Eisenstein2023,
       author = {{Eisenstein}, Daniel J. and {Willott}, Chris and {Alberts}, Stacey and {Arribas}, Santiago and {Bonaventura}, Nina and {Bunker}, Andrew J. and {Cameron}, Alex J. and {Carniani}, Stefano and {Charlot}, Stephane and {Curtis-Lake}, Emma and {D'Eugenio}, Francesco and {Endsley}, Ryan and {Ferruit}, Pierre and {Giardino}, Giovanna and {Hainline}, Kevin and {Hausen}, Ryan and {Jakobsen}, Peter and {Johnson}, Benjamin D. and {Maiolino}, Roberto and {Rieke}, Marcia and {Rieke}, George and {Rix}, Hans-Walter and {Robertson}, Brant and {Stark}, Daniel P. and {Tacchella}, Sandro and {Williams}, Christina C. and {Willmer}, Christopher N.~A. and {Baker}, William M. and {Baum}, Stefi and {Bhatawdekar}, Rachana and {Boyett}, Kristan and {Chen}, Zuyi and {Chevallard}, Jacopo and {Circosta}, Chiara and {Curti}, Mirko and {Danhaive}, A. Lola and {DeCoursey}, Christa and {de Graaff}, Anna and {Dressler}, Alan and {Egami}, Eiichi and {Helton}, Jakob M. and {Hviding}, Raphael E. and {Ji}, Zhiyuan and {Jones}, Gareth C. and {Kumari}, Nimisha and {L{\"u}tzgendorf}, Nora and {Laseter}, Isaac and {Looser}, Tobias J. and {Lyu}, Jianwei and {Maseda}, Michael V. and {Nelson}, Erica and {Parlanti}, Eleonora and {Perna}, Michele and {Pusk{\'a}s}, D{\'a}vid and {Rawle}, Tim and {Rodr{\'\i}guez Del Pino}, Bruno and {Sandles}, Lester and {Saxena}, Aayush and {Scholtz}, Jan and {Sharpe}, Katherine and {Shivaei}, Irene and {Silcock}, Maddie S. and {Simmonds}, Charlotte and {Skarbinski}, Maya and {Smit}, Renske and {Stone}, Meredith and {Suess}, Katherine A. and {Sun}, Fengwu and {Tang}, Mengtao and {Topping}, Michael W. and {{\"U}bler}, Hannah and {Villanueva}, Natalia C. and {Wallace}, Imaan E.~B. and {Whitler}, Lily and {Witstok}, Joris and {Woodrum}, Charity},
        title = "{Overview of the JWST Advanced Deep Extragalactic Survey (JADES)}",
      journal = {arXiv e-prints},
         year = 2023,
        month = jun,
          eid = {arXiv:2306.02465},
        pages = {arXiv:2306.02465},
          doi = {10.48550/arXiv.2306.02465},
archivePrefix = {arXiv},
       eprint = {2306.02465},
 primaryClass = {astro-ph.GA},
       adsurl = {https://ui.adsabs.harvard.edu/abs/2023arXiv230602465E}
}

@ARTICLE{Illingworth2016,
       author = {{Illingworth}, Garth and {Magee}, Daniel and {Bouwens}, Rychard and {Oesch}, Pascal and {Labbe}, Ivo and {van Dokkum}, Pieter and {Whitaker}, Katherine and {Holden}, Bradford and {Franx}, Marijn and {Gonzalez}, Valentino},
        title = "{The Hubble Legacy Fields (HLF-GOODS-S) v1.5 Data Products: Combining 2442 Orbits of GOODS-S/CDF-S Region ACS and WFC3/IR Images}",
      journal = {arXiv e-prints},
         year = 2016,
        month = jun,
          eid = {arXiv:1606.00841},
        pages = {arXiv:1606.00841},
          doi = {10.48550/arXiv.1606.00841},
archivePrefix = {arXiv},
       eprint = {1606.00841},
 primaryClass = {astro-ph.GA},
       adsurl = {https://ui.adsabs.harvard.edu/abs/2016arXiv160600841I}
}

@ARTICLE{Fontana2000,
       author = {{Fontana}, Adriano and {D'Odorico}, Sandro and {Poli}, Francesco and {Giallongo}, Emanuele and {Arnouts}, Stephane and {Cristiani}, Stefano and {Moorwood}, Alan and {Saracco}, Paolo},
        title = "{Photometric Redshifts and Selection of High-Redshift Galaxies in the NTT and Hubble Deep Fields}",
      journal = {\aj},
         year = 2000,
        month = nov,
       volume = {120},
       number = {5},
        pages = {2206-2219},
          doi = {10.1086/316803},
archivePrefix = {arXiv},
       eprint = {astro-ph/0009158},
 primaryClass = {astro-ph},
       adsurl = {https://ui.adsabs.harvard.edu/abs/2000AJ....120.2206F}
}

@ARTICLE{Brammer2008,
       author = {{Brammer}, Gabriel B. and {van Dokkum}, Pieter G. and {Coppi}, Paolo},
        title = "{EAZY: A Fast, Public Photometric Redshift Code}",
      journal = {\apj},
         year = 2008,
        month = oct,
       volume = {686},
       number = {2},
        pages = {1503-1513},
          doi = {10.1086/591786},
archivePrefix = {arXiv},
       eprint = {0807.1533},
 primaryClass = {astro-ph},
       adsurl = {https://ui.adsabs.harvard.edu/abs/2008ApJ...686.1503B}
}

@ARTICLE{Walcher2011,
       author = {{Walcher}, Jakob and {Groves}, Brent and {Budav{\'a}ri}, Tam{\'a}s and {Dale}, Daniel},
        title = "{Fitting the integrated spectral energy distributions of galaxies}",
      journal = {\apss},
         year = 2011,
        month = jan,
       volume = {331},
       number = {1},
        pages = {1-51},
          doi = {10.1007/s10509-010-0458-z},
archivePrefix = {arXiv},
       eprint = {1008.0395},
 primaryClass = {astro-ph.CO},
       adsurl = {https://ui.adsabs.harvard.edu/abs/2011Ap&SS.331....1W}
}

@ARTICLE{Pozzetti2010,
       author = {{Pozzetti}, L. and {Bolzonella}, M. and {Zucca}, E. and {Zamorani}, G. and {Lilly}, S. and {Renzini}, A. and {Moresco}, M. and {Mignoli}, M. and {Cassata}, P. and {Tasca}, L. and {Lamareille}, F. and {Maier}, C. and {Meneux}, B. and {Halliday}, C. and {Oesch}, P. and {Vergani}, D. and {Caputi}, K. and {Kova{\v{c}}}, K. and {Cimatti}, A. and {Cucciati}, O. and {Iovino}, A. and {Peng}, Y. and {Carollo}, M. and {Contini}, T. and {Kneib}, J. -P. and {Le F{\'e}vre}, O. and {Mainieri}, V. and {Scodeggio}, M. and {Bardelli}, S. and {Bongiorno}, A. and {Coppa}, G. and {de la Torre}, S. and {de Ravel}, L. and {Franzetti}, P. and {Garilli}, B. and {Kampczyk}, P. and {Knobel}, C. and {Le Borgne}, J. -F. and {Le Brun}, V. and {Pell{\`o}}, R. and {Perez Montero}, E. and {Ricciardelli}, E. and {Silverman}, J.~D. and {Tanaka}, M. and {Tresse}, L. and {Abbas}, U. and {Bottini}, D. and {Cappi}, A. and {Guzzo}, L. and {Koekemoer}, A.~M. and {Leauthaud}, A. and {Maccagni}, D. and {Marinoni}, C. and {McCracken}, H.~J. and {Memeo}, P. and {Porciani}, C. and {Scaramella}, R. and {Scarlata}, C. and {Scoville}, N.},
        title = "{zCOSMOS - 10k-bright spectroscopic sample. The bimodality in the galaxy stellar mass function: exploring its evolution with redshift}",
      journal = {\aap},
         year = 2010,
        month = nov,
       volume = {523},
          eid = {A13},
        pages = {A13},
          doi = {10.1051/0004-6361/200913020},
archivePrefix = {arXiv},
       eprint = {0907.5416},
 primaryClass = {astro-ph.CO},
       adsurl = {https://ui.adsabs.harvard.edu/abs/2010A&A...523A..13P}
}

@misc{Bradley2024,
       author = {{Bradley}, Larry and {Sip{\H{o}}cz}, Brigitta and {Robitaille}, Thomas and {Tollerud}, Erik and {Vin{\'\i}cius}, Z{\'e} and {Deil}, Christoph and {Barbary}, Kyle and {Wilson}, Tom J and {Busko}, Ivo and {Donath}, Axel and {G{\"u}nther}, Hans Moritz and {Cara}, Mihai and {Lim}, P.~L. and {Me{\ss}linger}, Sebastian and {Conseil}, Simon and {Burnett}, Zach and {Bostroem}, Azalee and {Droettboom}, Michael and {Bray}, E.~M. and {Andersen Bratholm}, Lars and {Ginsburg}, Adam and {Jamieson}, William and {Barentsen}, Geert and {Craig}, Matt and {Morris}, Brett M. and {Perrin}, Marshall and {Rathi}, Shivangee and {Pascual}, Sergio and {Georgiev}, Iskren Y.},
        title = "{astropy/photutils: 2.0.2}",
         year = 2024,
        month = oct,
          eid = {10.5281/zenodo.13989456},
          doi = {10.5281/zenodo.13989456},
      version = {2.0.2},
    publisher = {Zenodo},
       adsurl = {https://ui.adsabs.harvard.edu/abs/2024zndo..13989456B}
}

@ARTICLE{Matsuura2024,
       author = {{Matsuura}, Mikako and {Boyer}, M. and {Arendt}, Richard G. and {Larsson}, J. and {Fransson}, C. and {Rest}, A. and {Ravi}, A.~P. and {Park}, S. and {Cigan}, P. and {Temim}, T. and {Dwek}, E. and {Barlow}, M.~J. and {Bouchet}, P. and {Clayton}, G. and {Chevalier}, R. and {Danziger}, J. and {De Buizer}, J. and {De Looze}, I. and {De Marchi}, G. and {Fox}, O. and {Gall}, C. and {Gehrz}, R.~D. and {Gomez}, H.~L. and {Indebetouw}, R. and {Kangas}, T. and {Kirchschlager}, F. and {Kirshner}, R. and {Lundqvist}, P. and {Marcaide}, J.~M. and {Mart{\'\i}-Vidal}, I. and {Meixner}, M. and {Milisavljevic}, D. and {Orlando}, S. and {Otsuka}, M. and {Priestley}, F. and {Richards}, A.~M.~S. and {Schmidt}, F. and {Staveley-Smith}, L. and {Smith}, Nathan and {Spyromilio}, J. and {Vink}, J. and {Wang}, Lifan and {Watson}, D. and {Wesson}, R. and {Wheeler}, J.~C. and {Woodward}, C.~E. and {Zanardo}, G. and {Alp}, D. and {Burrows}, D.},
        title = "{Deep JWST/NIRCam imaging of Supernova 1987A}",
      journal = {\mnras},
         year = 2024,
        month = aug,
       volume = {532},
       number = {4},
        pages = {3625-3642},
          doi = {10.1093/mnras/stae1032},
archivePrefix = {arXiv},
       eprint = {2404.10042},
 primaryClass = {astro-ph.SR},
       adsurl = {https://ui.adsabs.harvard.edu/abs/2024MNRAS.532.3625M}
}

@ARTICLE{Berkheimer2024,
       author = {{Berkheimer}, Jessica M. and {Carleton}, Timothy and {Windhorst}, Rogier A. and {Keel}, William C. and {Holwerda}, Benne W. and {Nonino}, Mario and {Cohen}, Seth H. and {Jansen}, Rolf A. and {Coe}, Dan and {Conselice}, Christopher J. and {Driver}, Simon P. and {Frye}, Brenda L. and {Grogin}, Norman A. and {Koekemoer}, Anton M. and {Lucas}, Ray A. and {Marshall}, Madeline A. and {Pirzkal}, Nor and {Robertson}, Clayton and {Robotham}, Aaron and {Ryan}, Russell E. and {Smith}, Brent M. and {Summers}, Jake and {Tompkins}, Scott and {Willmer}, Christopher N.~A. and {Yan}, Haojing},
        title = "{JWST NIRCam Photometry: A Study of Globular Clusters Surrounding Bright Elliptical Galaxy VV 191a at z = 0.0513}",
      journal = {\apjl},
         year = 2024,
        month = apr,
       volume = {964},
       number = {2},
          eid = {L29},
        pages = {L29},
          doi = {10.3847/2041-8213/ad2688},
archivePrefix = {arXiv},
       eprint = {2310.16923},
 primaryClass = {astro-ph.GA},
       adsurl = {https://ui.adsabs.harvard.edu/abs/2024ApJ...964L..29B}
}

@ARTICLE{Maraston2010,
       author = {{Maraston}, Claudia and {Pforr}, Janine and {Renzini}, Alvio and {Daddi}, Emanuele and {Dickinson}, Mark and {Cimatti}, Andrea and {Tonini}, Chiara},
        title = "{Star formation rates and masses of z \raisebox{-0.5ex}\textasciitilde 2 galaxies from multicolour photometry}",
      journal = {\mnras},
         year = 2010,
        month = sep,
       volume = {407},
       number = {2},
        pages = {830-845},
          doi = {10.1111/j.1365-2966.2010.16973.x},
archivePrefix = {arXiv},
       eprint = {1004.4546},
 primaryClass = {astro-ph.CO},
       adsurl = {https://ui.adsabs.harvard.edu/abs/2010MNRAS.407..830M}
}

@ARTICLE{Boquien2010,
       author = {{Boquien}, M. and {Duc}, P. -A. and {Galliano}, F. and {Braine}, J. and {Lisenfeld}, U. and {Charmandaris}, V. and {Appleton}, P.~N.},
        title = "{Star Formation in Collision Debris: Insights from the Modeling of Their Spectral Energy Distribution}",
      journal = {\aj},
         year = 2010,
        month = dec,
       volume = {140},
       number = {6},
        pages = {2124-2144},
          doi = {10.1088/0004-6256/140/6/2124},
archivePrefix = {arXiv},
       eprint = {1010.2201},
 primaryClass = {astro-ph.CO},
       adsurl = {https://ui.adsabs.harvard.edu/abs/2010AJ....140.2124B}
}

@ARTICLE{deBarros2014,
       author = {{de Barros}, S. and {Schaerer}, D. and {Stark}, D.~P.},
        title = "{Properties of z \raisebox{-0.5ex}\textasciitilde 3-6 Lyman break galaxies. II. Impact of nebular emission at high redshift}",
      journal = {\aap},
         year = 2014,
        month = mar,
       volume = {563},
          eid = {A81},
        pages = {A81},
          doi = {10.1051/0004-6361/201220026},
archivePrefix = {arXiv},
       eprint = {1207.3663},
 primaryClass = {astro-ph.CO},
       adsurl = {https://ui.adsabs.harvard.edu/abs/2014A&A...563A..81D}
}

@ARTICLE{Wuyts2012,
       author = {{Wuyts}, Stijn and {F{\"o}rster Schreiber}, Natascha M. and {Genzel}, Reinhard and {Guo}, Yicheng and {Barro}, Guillermo and {Bell}, Eric F. and {Dekel}, Avishai and {Faber}, Sandra M. and {Ferguson}, Henry C. and {Giavalisco}, Mauro and {Grogin}, Norman A. and {Hathi}, Nimish P. and {Huang}, Kuang-Han and {Kocevski}, Dale D. and {Koekemoer}, Anton M. and {Koo}, David C. and {Lotz}, Jennifer and {Lutz}, Dieter and {McGrath}, Elizabeth and {Newman}, Jeffrey A. and {Rosario}, David and {Saintonge}, Amelie and {Tacconi}, Linda J. and {Weiner}, Benjamin J. and {van der Wel}, Arjen},
        title = "{Smooth(er) Stellar Mass Maps in CANDELS: Constraints on the Longevity of Clumps in High-redshift Star-forming Galaxies}",
      journal = {\apj},
         year = 2012,
        month = jul,
       volume = {753},
       number = {2},
          eid = {114},
        pages = {114},
          doi = {10.1088/0004-637X/753/2/114},
archivePrefix = {arXiv},
       eprint = {1203.2611},
 primaryClass = {astro-ph.CO},
       adsurl = {https://ui.adsabs.harvard.edu/abs/2012ApJ...753..114W}
}

@ARTICLE{Osborne2024,
       author = {{Osborne}, Chandler and {Salim}, Samir},
        title = "{Strategies for Obtaining Robust Spectral Energy Distribution Fitting Parameters for Galaxies at z {\ensuremath{\sim}} 1 and z {\ensuremath{\sim}} 2 in the Absence of Infrared Data}",
      journal = {\apj},
         year = 2024,
        month = feb,
       volume = {962},
       number = {1},
          eid = {59},
        pages = {59},
          doi = {10.3847/1538-4357/ad17c8},
archivePrefix = {arXiv},
       eprint = {2401.06865},
 primaryClass = {astro-ph.GA},
       adsurl = {https://ui.adsabs.harvard.edu/abs/2024ApJ...962...59O}
}

@ARTICLE{Inoue2011,
       author = {{Inoue}, Akio K.},
        title = "{Rest-frame ultraviolet-to-optical spectral characteristics of extremely metal-poor and metal-free galaxies}",
      journal = {\mnras},
         year = 2011,
        month = aug,
       volume = {415},
       number = {3},
        pages = {2920-2931},
          doi = {10.1111/j.1365-2966.2011.18906.x},
archivePrefix = {arXiv},
       eprint = {1102.5150},
 primaryClass = {astro-ph.CO},
       adsurl = {https://ui.adsabs.harvard.edu/abs/2011MNRAS.415.2920I}
}

\FloatBarrier
\clearpage
\begin{appendix}
\section{SFH parametrization}\label{app_SFH}
To explore the impact of SFH assumptions on our results, we tested several commonly used parametric SFH models, including constant, delayed, and delayed with an optional recent burst in the last 100~Myr. Our goal was to assess whether the choice of SFH significantly affects the stellar mass differences between resolved and integrated SED fits. For each SFH scenario (with parameters outlined in Table~\ref{tab:sfhparam}), we performed SED fitting using the same dataset and configuration, and compared the resulting $\Delta \log \mathrm{M}_\star$ values. We also tested variations in the minimum stellar population age allowed for the stellar population, including cutoffs at 10, 25, 50, and 100~Myr, in order to test its consequence on amplifying the outshining bias.

\begin{figure}[h]
\centering
\includegraphics[width=0.5\textwidth]{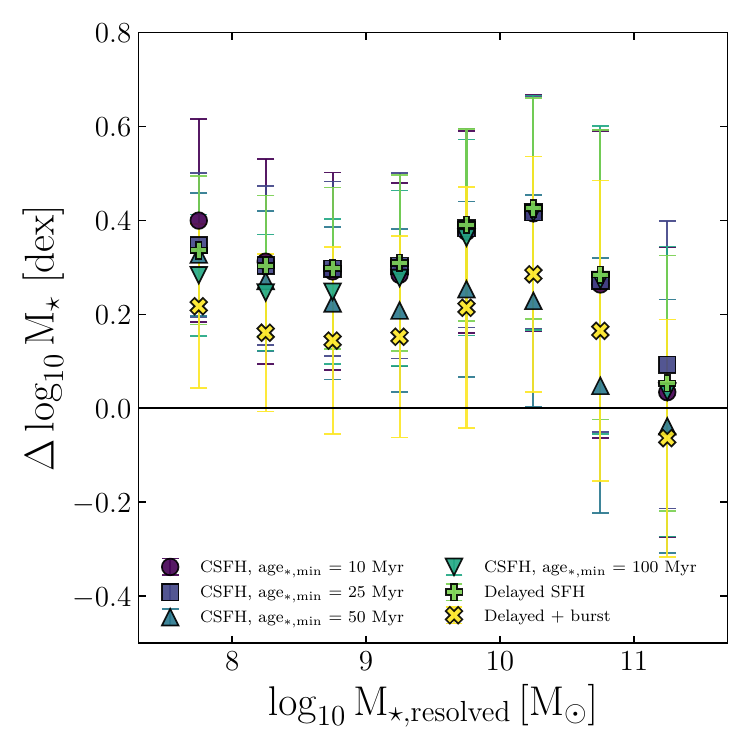}
\caption{Difference between resolved and integrated stellar masses as a function of resolved stellar mass, for different SFH assumptions. Error bars indicate the 16th-84th percentile range in each bin. Results are shown for constant SFHs (CSFH) with minimum stellar ages of 10, 25, 50, and 100~Myr, delayed SFH, and delayed + burst SFH.}
\label{Fig.SFH}
\end{figure}

We found that the overall trends and mass discrepancies between resolved and integrated fits were broadly consistent across these choices. The median values of $\Delta \log \mathrm{M}_\star$ for the constant SFH using cutoffs of 10 and 25~Myr were among the highest, at 0.29 and 0.30~dex respectively, while the delayed SFH also yielded a similarly high offset of 0.30~dex. The lowest offset was obtained with the delayed + burst model (0.16~dex). The difference between the highest and lowest median $\Delta \log \mathrm{M}_\star$ across all configurations is 0.14~dex. We adopted a constant SFH with a minimum age of 50~Myr for our final modeling, as it yielded an intermediate offset (0.24~dex) and the lowest scatter among all tested configurations ($\sigma = 0.18$~dex, compared to $\sigma = 0.21$~dex for constant with 10~Myr cutoff and $\sigma = 0.23$~dex for delayed + burst). Additionally, the constant SFH has fewer free parameters than delayed or burst models, making it computationally efficient, which was a practical consideration given the large number of pixels fitted in our resolved analysis. This choice also helped mitigate possible outshining biases by avoiding the need to populate galaxies with very young stellar ages, which can dominate the light. These results are illustrated in Figure~\ref{Fig.SFH}. This finding is supported by spectro-photometric studies which demonstrate that stellar mass estimates are robust to variations in SFH assumptions \citep{Annunziatella2025}.

\begin{table}[]
\caption{Summary of input parameters for the various SFH scenarios tested in the SED fitting using {\ttfamily CIGALE}. Each SFH model was applied independently to assess the robustness of our results. All other components of the SED modeling, such as dust attenuation, nebular emission, and stellar population synthesis, were held fixed and consistent with the configuration listed in Table~\ref{tab:sedparam}. The values were linearly spaced.}\label{tab:sfhparam}
  \begin{center}
    \begin{tabular}{l r}
     \hline\hline
    \textbf{Parameter}\ \ \ \ \ \ \ \ \ \ \ \ \ \ \ \ \ \ \ \ \ \ \ \ & \ \ \ \ \ \ \ \ \ \ \ \ \ \ \ \ \ \ \ \ \ \ \ \ \textbf{Priors}\\
     \hline\hline
     \multicolumn{2}{c}{Constant SFH}   \\
     \hline
    age$_{\star}$ [Myr]  & 100 values in [10-2000]\\
    
      & 100 values in [25-2000]\\
    
     & 100 values in [50-2000]\\
    
      & 100 values in [100-2000]\\
    \hline\hline
    \multicolumn{2}{c}{Delayed SFH}   \\
     \hline
     $\tau$$^\mathrm{i}$ [Myr] & 20 values in [100-6000]\\
     age$_{\star}$ [Myr]  & 100 values in [50-2000]\\
    \hline\hline
    \multicolumn{2}{c}{Delayed SFH + recent burst}   \\
     \hline
     $\tau_{\rm{main}}$ [Myr] & 20 values in [100-6000]\\
     $\tau_{\rm{burst}}$$^{\mathrm{ii}}$ [Myr]  & 5 values in [50-500]\\
     $f_{\rm{burst}}$$^{\mathrm{iii}}$ &  0$\%$, 1$\%$, 5$\%$, 10$\%$\\
     age$_{\star}$  [Myr] & 100 values in [50-2000]\\
     age$_{\rm{burst}}$ [Myr] & 10, 25, 50\\
    \hline\hline
    \end{tabular}
    \end{center}
    $^\mathrm{i}$ e-folding time of the main stellar population.\\
    $^{\mathrm{ii}}$ e-folding time of the late stellar population.\\
    $^{\mathrm{iii}}$ Mass fraction of the late burst population.\\
    \end{table}

\section{Comparisons between SED fitting approaches}\label{SED_comparison}
\begin{figure*}[]
\centering
\includegraphics[width=1\textwidth]{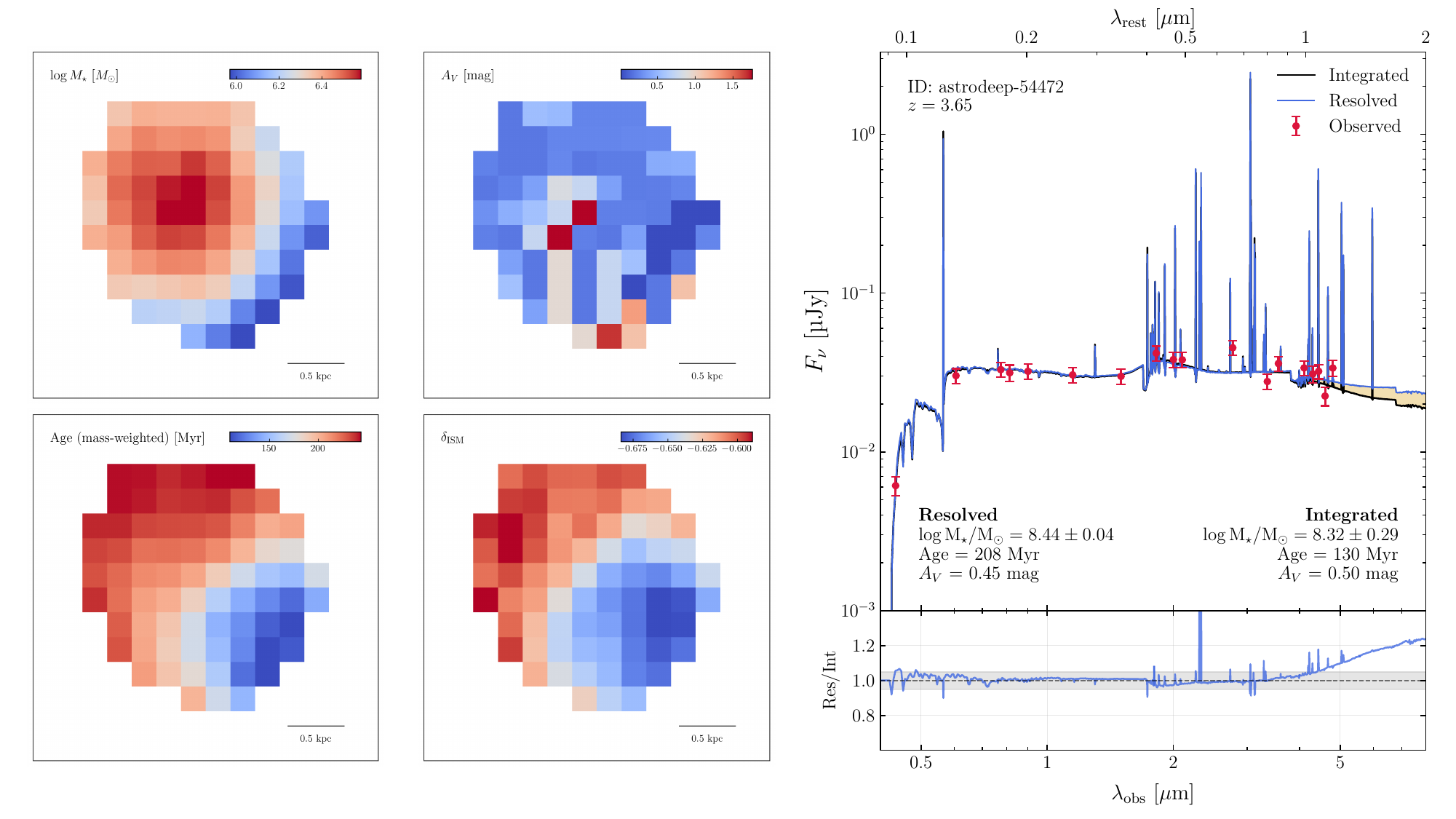}
\caption{Resolved properties and SED comparison for galaxy astrodeep-54472 
at $z = 3.65$. \textbf{Left}: Spatially resolved maps of stellar mass, 
$A_V$, mass-weighted age, and $\delta_{\mathrm{ISM}}$ from pixel-by-pixel 
SED fitting, revealing strong central concentrations of mass and attenuation, age gradients, and spatial variations in attenuation slope. \textbf{Right (top)}: Observed photometry (red points) compared to integrated (black) and summed resolved (blue) SED models. The gold shaded region shows flux differences between methods. Resolved fitting recovers higher stellar mass (8.44 vs 8.32 dex) and older age (208 vs 130~Myr). \textbf{Right (bottom)}: Ratio of resolved to integrated SEDs (Res/Int); dashed line marks unity, gray band shows $\pm5\%$. This example illustrates how spatial gradients in dust and stellar populations, visible in resolved analysis, are averaged out in integrated fits, leading to systematically lower mass and age estimates.}
\label{Fig.SED_comparison}
\end{figure*}
To assess how different measurement approaches influence the derived stellar properties, we applied both resolved and integrated SED analyses to each galaxy in our sample. In the integrated approach, total fluxes were obtained by summing the pixels that satisfy our predefined S/N and band coverage thresholds (Section~\ref{sample}), ensuring consistency with the resolved analysis. The resolved approach was constructed by summing the best-fit pixel-by-pixel SEDs, which were then compared directly to the integrated SEDs for each galaxy.

For reference, we also considered aperture photometry from \citet{Merlin2024}, corresponding to standard one-dimensional fitting applied to the total flux measured within a large aperture encompassing the full galaxy. The stellar masses derived from both the resolved and integrated analyses were scaled using aperture-based corrections to ensure that all measurements refer to the same total enclosed flux.

Figure~\ref{Fig.SED_comparison} illustrates the comparison between these two approaches for a representative galaxy. The spatially resolved property maps (left panels) reveal strong internal gradients: stellar mass and dust 
attenuation ($A_V$) are centrally concentrated, mass-weighted ages increase 
toward the galaxy center, and the ISM attenuation slope ($\delta_{\mathrm{ISM}}$) varies spatially across the galaxy. These gradients demonstrate the internal heterogeneity that is averaged over in integrated fitting. Both the resolved and integrated approaches (right panels) provide good fits to the observed photometry, but yield systematically different physical parameters. The resolved approach, which utilizes spatial information at the pixel level, recovers a higher stellar mass (0.12~dex) and older mass-weighted age ($\sim$60\% older) compared to the integrated approach that treats the galaxy as a single unresolved source. This example demonstrates the key systematic differences that persist across our full sample, which are quantified statistically in Section~\ref{results}.

\section{Mock analysis}\label{app_mock}
To evaluate the reliability of the derived physical parameters, we performed a mock analysis using the best-fit SEDs from our main fitting run. Specifically, for each galaxy, synthetic photometric fluxes were generated based on the best-fit model in the observed bands and then perturbed with random Gaussian noise consistent with the measurement uncertainties in the original catalog. This mock photometry was subsequently refitted using the same configuration as the original analysis. By comparing the input (exact) and recovered (estimated) physical parameters, such as stellar mass, SFR, and attenuation, we obtain an empirical estimate of the typical uncertainties and potential biases associated with the adopted SED modeling setup. This provides a crucial test of the internal consistency and robustness of our fitting approach \citep{Boquien2019, Osborne2024}.\\

\begin{figure*}[]
\centering
\includegraphics[width=1\textwidth]{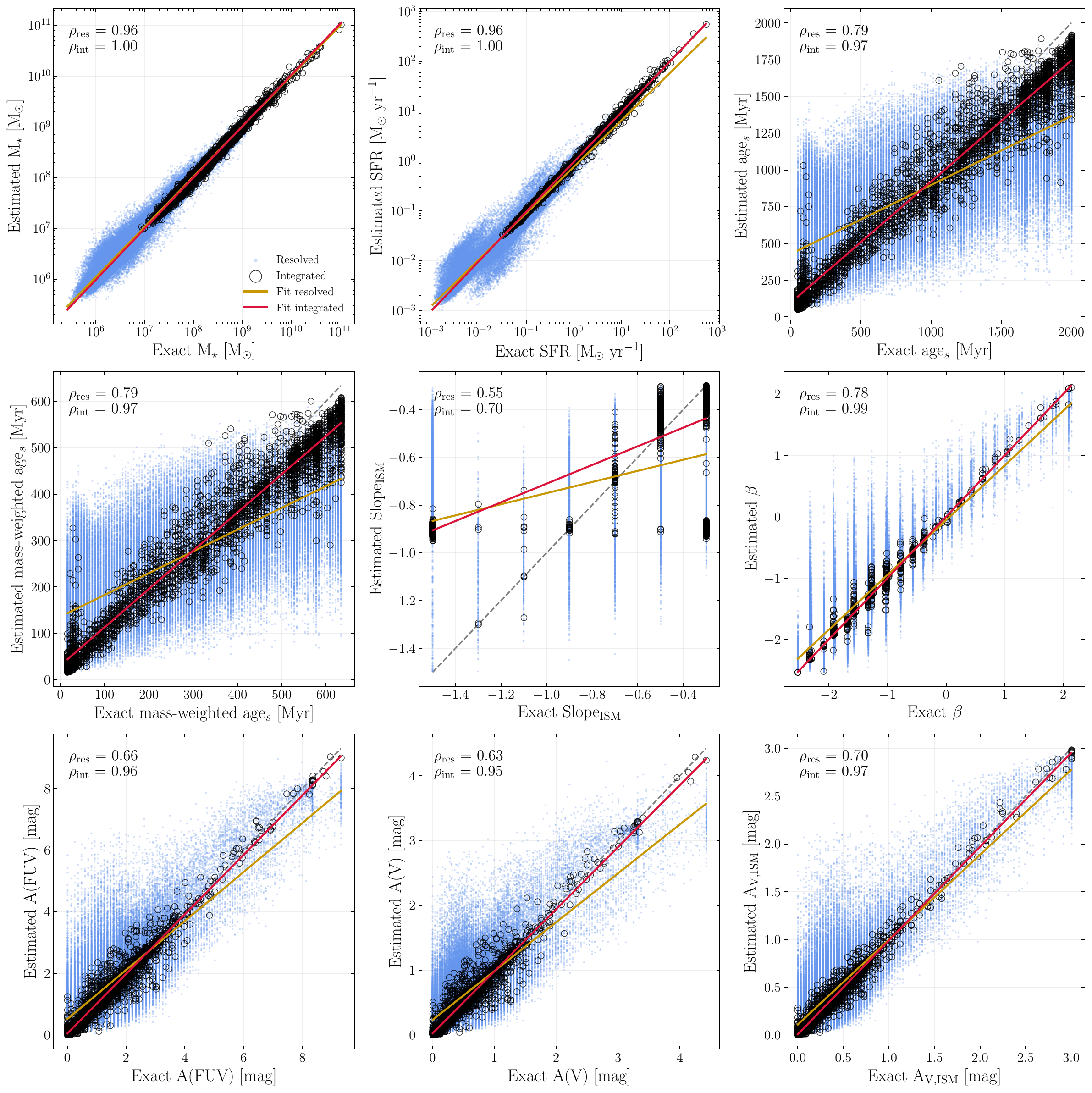}
\caption{Comparison between the exact input values (x-axis) and the recovered mock values (y-axis) derived from the mock analysis of the resolved (blue crosses) and integrated (yellow dots). Each point represents the probability weighted mean of the posterior distribution for a given parameter. The stellar mass, SFR, and the $\beta$ UV slope are all a result of SED fitting. The other parameters, namely the ages, the attenuation slopes and the attenuation values in the ISM are all priors for the SED fitting process. The black dashed line indicates the one-to-one relation. The red and blue solid lines represent linear fits to the integrated and resolved results, respectively. Spearman's rank correlation coefficients ($\rho$) for each fit are indicated in the legend of each panel.}
\label{Fig.mock}
\end{figure*}

The results of the mock analysis are illustrated in Figure \ref{Fig.mock}. Each panel compares a key physical parameter (e.g., stellar mass, SFR, age, dust attenuation properties, and the attenuation slope) estimated using both integrated and spatially resolved SED fitting. Overall, most parameters show strong correlations with their exact input values, particularly stellar mass and SFR, which are recovered with high fidelity ($\rho \approx 0.99$ for integrated and ($\rho \approx 0.93$ for resolved fits). Some parameters, such as the attenuation slope and ISM attenuation exhibit larger scatter, especially in the resolved case. This is expected, as these prior parameters are more sensitive to local variations in S/N and spectral coverage, and may suffer from degeneracies with age and SFH shape. Nonetheless, the recovered values of the stellar mass and SFR follow the one-to-one relation, indicating that our SED fitting setup is capable of reproducing the input parameters across a wide range of physical conditions. This validates the robustness of our modeling framework for both integrated and resolved analyses, while also rendering the computations less expensive.

\end{appendix}

\end{document}